\documentclass[twocolumn]{aastex631}    

\usepackage{booktabs}
\usepackage{longtable}
\usepackage{appendix}
\usepackage{amsmath}
\usepackage{amssymb}
\usepackage{graphicx}
\usepackage{mfirstuc}
\usepackage{rotating}
\usepackage{soul}

\newcommand{\hei}{He\,\textsc{i}}
\newcommand{\heii}{He\,\textsc{ii}}
\newcommand{\trans}[3]{\capitalisewords{#1}\,\textsc{#2}\,$\lambda$#3}

\graphicspath{{figures/, tables/}}
\shorttitle{Metal-Poor OB Stars in NGC~3109}
\shortauthors{Mintz et al.}

\begin{document}

\newcommand{\princeton}{Department of Astrophysical Sciences, Princeton University, 4 Ivy Lane, Princeton, NJ 08544, USA; {\color{xlinkcolor}abby.mintz@princeton.edu}} \newcommand{\carnegie}{The Observatories of the Carnegie Institution for Science, 813 Santa Barbara Street, Pasadena, CA 91101, USA}
\newcommand{\rutgers}{Department of Physics and Astronomy, Rutgers University, 136 Frelinghuysen Road, Piscataway, NJ 08854, USA}
\newcommand{\utaustin}{Department of Astronomy, The University of Texas at Austin, 2515 Speedway, Stop C1400, Austin, TX 78712-1205, USA}
\newcommand{\stsci}{Space Telescope Science Institute, 3700 San Martin Drive, Baltimore, MD 21218, USA}
\newcommand{\notredame}{Department of Physics and Astronomy, University of Notre Dame, 225 Nieuwland Science Hall, Notre Dame, IN 46556, USA}

\title{A Spectroscopic Survey of Metal-Poor OB Stars in Local Dwarf Galaxy NGC 3109}

\author[0000-0002-9816-9300]{Abby Mintz}
\affil{\princeton}
\author[0000-0003-4122-7749]{O.\ Grace Telford}
\affiliation{\princeton}
\affiliation{\carnegie}
\affiliation{\rutgers}
\author[0000-0001-6196-5162]{Evan N.\ Kirby}
\affiliation{\notredame}
\author[0000-0002-0302-2577]{John Chisholm}
\affiliation{\utaustin}
\author[0000-0001-5538-2614]{Kristen B.\ W.\ McQuinn}
\affiliation{\rutgers}
\affiliation{\stsci}
\author[0000-0002-4153-053X]{Danielle A.\ Berg}
\affiliation{\utaustin}

\begin{abstract}
As JWST uncovers increasingly strong evidence that metal-poor, massive stars in early galaxies dominated reionization, observational constraints on the properties of such stars are more relevant than ever before. However, spectra of individual O- and B-type stars are rare at the relevant metallicities ($\lesssim 0.2$ $Z_\odot$), leaving models of stellar evolution and ionizing flux poorly constrained by data in this regime. We present new medium-resolution ($R\sim 4000)$ Keck/DEIMOS optical spectra of 17 OB stars in the local low-metallicity (0.12 $Z_\odot$) dwarf galaxy NGC 3109. We assign spectral types to the stars and present new criteria for selecting O stars using optical and NUV photometry from Hubble Space Telescope imaging. We fit the spectra and photometry with grids of stellar atmosphere models to measure stellar temperatures, surface gravities, luminosities, radii, and masses. We find evidence of strong mass loss via radiation-driven stellar winds in two O stars, one of which is the hottest, youngest, and most massive star confirmed in the host galaxy to date. Though its spectrum does not meet conventional Wolf-Rayet spectral classification criteria, this metal-poor O~If star produces strong \trans{he}{ii}{4686} emission and its evolutionary status is ambiguous. This work nearly doubles the number of OB stars with measured parameters in NGC 3109, including ten stars with no previously reported parameters, four with no published spectroscopy, and four binary candidates. This large sample of OB stellar parameters provides a new observational testbed to constrain the stellar astrophysics that drove cosmic reionization and influenced the evolution of the earliest galaxies.
\end{abstract}


\section{Introduction} \label{sec:intro}

\begin{figure*}[t]
\begin{center}
\includegraphics[width=\linewidth,angle=0]{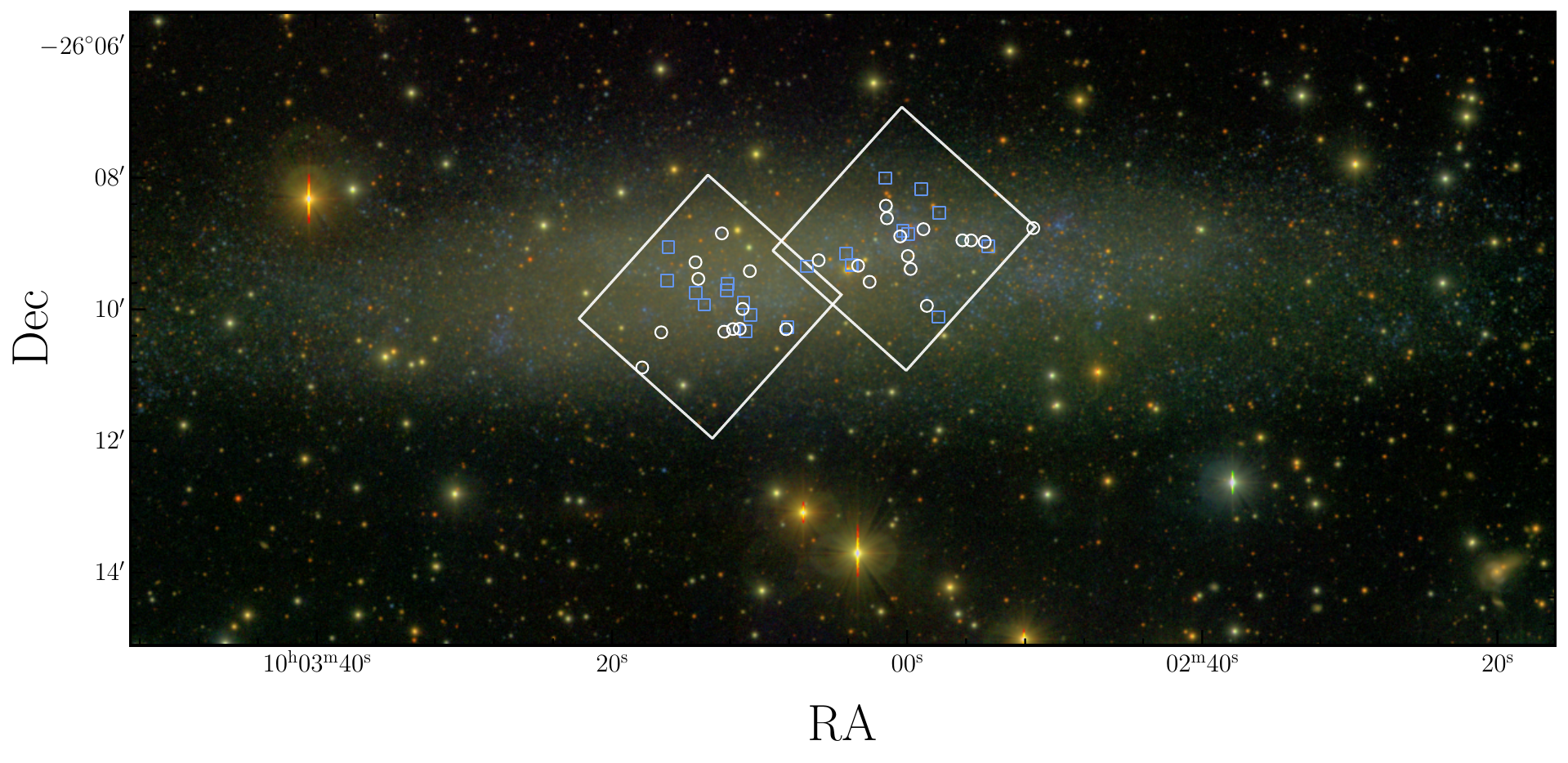}
\caption{Location of candidate OB stars in NGC 3109. The footprints of the HST imaging pointings are shown as white outlines. Candidate OB stars selected with HST photometry as described in Section~\ref{subsec:spec} are shown as white circles and blue squares. The white circles represent stars which we observed with Keck/DEIMOS and blue squares represent OB candidates which did not fit on the slitmask and were therefore not observed. The background image of NGC 3109 is made with optical (\textit{BVR}) images from the Spitzer Local Volume Legacy Survey \citep{Cook2014}. }\label{fig:hamap}
\end{center}
\end{figure*}

Massive stars are the least numerous component of stellar populations, but they play an integral role in regulating the properties of their host galaxies at all scales. During their lives, massive stars shape the interstellar medium with their ionizing radiation and strong metal-line driven mass loss \citep{Puls2008, ElBadry2016, Emerick2018}. These stars govern subsequent generations of star formation, enrich interstellar material, and undergo dramatic supernova explosions as they evolve into neutron stars and black holes \citep{Stinson2007, Langer2012, Eldridge2013, Eldridge2022}.

Translating observations of star-forming galaxies into estimates of their star-formation rates, stellar demographics, and cosmic histories requires a thorough understanding of the evolutionary tracks of these massive stars \citep{Leitherer1999, Eldridge2009, Conroy2013, Chisholm2019}. The blue regions of spectral energy distributions (SEDs) and nebular line emission in these galaxies are directly determined by their OB-star populations and evolution \citep{Byler2017}. While such models are well tested at Galactic metallicity and significant progress has been made toward building lower-metallicity samples in the Magallenic Clouds \citep[e.g,][]{Vink2023}, theoretical predictions are almost completely unconstrained for more metal-poor stars ($Z<20\%\,Z_\odot$). Mass-loss rates \citep{Leitherer1992, Vink2001,Krticka2018, Bjorklund2021, Vink2021,  GormazMatamala2024}, ionizing flux production \citep{Martins2021}, and evolutionary pathways \citep{Szecsi2015,Groh2019, Szecsi2022} for metal-poor stars are predicted by extrapolation to lower $Z$ than probed by large observational samples of individual OB stars. As a result, conclusions drawn from using these theoretical stellar models to interpret unresolved observations of low-metallicity galaxies are highly uncertain. Thorough evaluation of these models requires empirical measurements of mass-loss rates and fundamental stellar parameters -- effective temperature ($T_\textrm{eff}$), surface gravity ($\log g$), and rotation speed ($v\sin i$) -- for a large sample of metal-poor stars.

As JWST pushes the high-redshift frontier to increasingly early cosmic epochs and consequently draws more attention to metal-poor stars and galaxies, additional observational constraints on low-metallicity massive-star theory are more needed than ever before. The number of JWST-detected reionization-era galaxies is growing rapidly \citep[e.g.,][]{Roberts-Borsani2023, Mascia2023}, providing increasing evidence that metal-poor O stars, which dominate ionizing flux in star-forming galaxies, likely drove cosmic reionization \citep{Topping2022, Nanayakkara2023, Atek2024}. Additional spectroscopic observations of large samples of local analog stars are required to constrain the stellar population synthesis (SPS) models relied upon to interpret these observations and infer these galaxies' assembly histories and ionizing photon productions \citep{Tinsley1980, Conroy2013}. 

It is well established that stellar properties and evolution vary with metallicity \citep{Heger2003, Eldridge2004}, but the precise nature of this dependence is complex and largely theoretical. Massive stars at Galactic metallicity experience significant mass loss over the course of their life, but metal-poor OB stars drive much weaker winds and retain more mass, resulting in a shift in their evolutionary pathways as a function of metallicity \citep{Szecsi2022}. {A lower metal content also leads to lower atmospheric opacity, increasing both the luminosity and temperature of the star as governed by the Stefan-Boltzmann law} \citep{Kippenhahn2013, Choi2016}. Stellar populations at low \textit{Z} therefore occupy a different region of the Hertzsprung-Russell diagram. These effects are highly non-linear and so observational constraints, including measurements of stellar temperature and luminosity for massive samples, are crucial to inform theoretical predictions.

Dwarf galaxies, which have low gas-phase metallicities, are prime candidates for observing metal-poor massive stars. However, measurements of fundamental stellar properties ($T_\text{eff}, \log g, v\sin i$) require optical spectroscopy, which has historically limited this work to the Local Group where stars are resolved and sufficiently bright. In recent years, extensive work has been done to catalog, classify, and characterize the population of OB stars in the Large and Small Magellanic Clouds \citep[LMC, SMC;][]{Massey2004, Mokiem2006, Evans2006, Massey2009, Lamb2016, Ramachandran2019, Vink2023, Sana2024}, but these galaxies have metallicities of 50\%\,$Z_\odot$ and 20\%\,$Z_\odot$, respectively, and so cannot probe the lower metallicities that are representative of early-universe galaxies. 

To do so, we must look farther afield to more distant low-metallicity dwarf galaxies where spectroscopic observations of individual massive stars become even more difficult and expensive. While a number of studies have undertaken this challenge and observed O and B stars in local dwarfs, the total sample of massive metal-poor stars with well-measured stellar parameters remains small. Stellar parameters have been published for only a few dozen OB stars in the metal-poor galaxies WLM, IC 1613, and NGC 3109 and similar numbers for OB stars in extremely metal-poor ($Z<10\%\,Z_\odot$) galaxies Sextans A, Leo A, and Leo P \citep{Kaufer2004,  Bresolin2006, Bresolin2007, Evans2007, Tramper2011, Tramper2014, Hosek2014, Camacho2016, Telford2021, Gull2022, Telford2024}. Given that fewer than a hundred individual OB stars have been identified with $Z<20\%\,Z_\odot$, every discovered sub-SMC metallicity OB star is an important constraint on the stellar astrophysics of the early universe.

    The local star-forming dwarf galaxy NGC 3109 is an especially compelling target for increasing the number of well-characterized metal-poor massive stars. At a distance of 1.34 Mpc \citep{Jacobs2009, Anand2021}, it is at the edge of the Local Group, and so is among the closest star-forming dwarf galaxies to the Milky Way. It is a sub-SMC metallicity galaxy with a gas-phase oxygen abundance of $12+\log(\textrm{O/H})=7.76$ \citep{Marble2010}, corresponding to a metallicity of 0.12 $Z_\odot$ with respect to the solar oxygen abundance \citep{Asplund2021}. In fact, NGC 3109 is of particular relevance as three of its OB stars have been included in the Hubble Space Telescope (HST) director's discretionary program UV Legacy Library of Young Stars as Essential Standards \citep[ULLYSES;][]{Roman-Duval2020}. ULLYSES has built a large catalog of FUV spectroscopy of metal-poor massive stars, enabling the measurement of their mass-loss rates and stellar abundances. The program has observed hundreds of stars in the LMC and SMC, but only six stars in sub-SMC metallicity galaxies -- three each in NGC 3109 and Sextans A.

    Only several dozen OB stars have been classified in NGC 3109 using low-resolution optical spectroscopy, and few of these stars have published stellar parameters. \citet{Evans2007} presented the first spectroscopic observations of resolved stars in NGC 3109, including only 12 late O stars among their 91 targets. However, the spectra were low resolution ($R\sim1000$) -- making it difficult to accurately measure $\log g$ and $T_\text{eff}$, which are constrained by the wings of the Balmer lines -- and \citet{Evans2007} measured stellar parameters for only eight B stars in their sample. \citet{Tramper2011} and \citet{Tramper2014} later measured stellar parameters of four of the previously classified O stars using medium-resolution spectra ($R\sim7000$) and \citet{Hosek2014} reanalyzed a number of late-B and A-type spectra from \citet{Evans2007}. No additional classifications or parameter measurements have been published for massive stars in NGC 3109 since these studies. 

    In this paper, we present a large sample of medium-resolution optical spectroscopy of OB stars in NGC 3109 selected based on new near-ultraviolet (NUV) and optical HST imaging from ULLYSES. The addition of NUV photometry provides an exciting opportunity to prioritize the hottest O stars, which are less easily distinguishable with optical photometry alone. Our sample includes stars classified by \citet{Evans2007} and stars with no previously published spectroscopy. We use our new spectroscopy to classify the stars and estimate their fundamental properties by fitting their spectroscopy and photometry with stellar atmosphere models. We present a catalog of the parameters we derive for 15 of the stars, among the largest set of OB-stellar parameters published for a sub-SMC galaxy. By nearly doubling the number of well-characterized massive stars in this metal poor galaxy, this work positions NGC~3109 as an ideal laboratory to study the demographics and properties of the low-metallicity stellar populations shaping the early universe.

    The structure of this paper is as follows. In Section~\ref{sec:data} we describe the HST photometry and the Keck/DEIMOS observations, including the selection procedure for the O-star candidates. In Section~\ref{sec:spectralclass}, we assign spectral types and luminosity classes (LC) to the stars, including five stars that have no previously published classifications. In Section~\ref{sec:fitting}, we describe the procedure to measure the stellar parameters by fitting model spectra and SEDs to our observed spectroscopy and photometry and we identify binary candidates. In Section~\ref{sec:res} we present our main results, including the identification of the youngest star so far detected in NGC 3109, which is among the most massive stars detected in any sub-SMC metallicity galaxy. In Section~\ref{sec:disc} we discuss the implications of our results, including the surprising detection of two metal-poor stars with signs of strong mass loss.

\section{Observations and Data Reduction} \label{sec:data}

\subsection{HST Photometry}
\begin{figure}[t]
\begin{center}
\includegraphics[width=1\linewidth,angle=0]{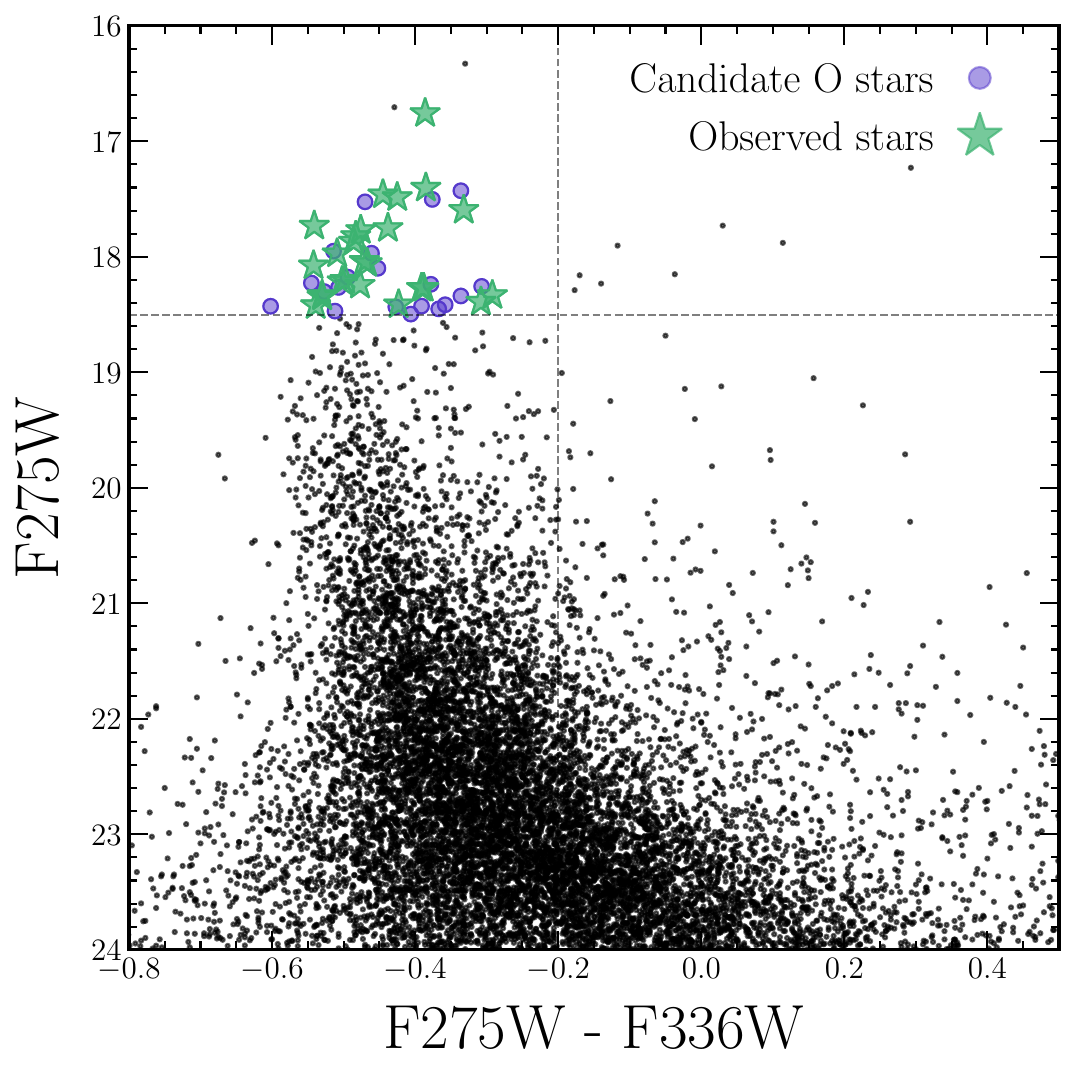}
\caption{The NUV color-magnitude diagram used in the selection of candidate O stars in NGC 3109. The black points show the F275W magnitude against the F275W -- F336W color for all stars in the two HST fields. The dashed lines indicate the criteria for O-star candidate selection: {F275W $<18.5$ mag} and F275W -- F336W $<-0.2$ mag. The candidate stars we did not observe are plotted as purple circles and the candidates which we observed with Keck/DEIMOS are shown as green stars. We note that the two brightest stars in the CMD were excluded because they were previously classified as B supergiants by \citet{Evans2007}. We separate the stars by spectral type in color-magnitude diagrams in Section~\ref{subsec:select} and \autoref{fig:big_cmd}. \label{fig:selection_cmd}}
\end{center}
\end{figure}

To identify bright and blue stars at the top of the main sequence in NGC~3109, we used 
NUV and optical HST imaging of two fields in this galaxy recently obtained by the ULLYSES program\footnote{The HST data analyzed in this paper can be downloaded from MAST at DOI\dataset[10.17909/gfe4-4230]{http://dx.doi.org/10.17909/gfe4-4230}}
 (HST-GO-16104; the two HST pointings are shown as white outlines in Figure~\ref{fig:hamap}).
Images were taken with UV/Visible (UVIS) channel of the Wide Field Camera 3 (WFC3) in 5 filters: F225W, F275W, F336W, F475W, and F814W\@. 
Here, we make use of this high-resolution imaging of  NGC~3109, particularly in the NUV filters, to search for O-type stars that may have been missed in previous surveys using ground-based optical photometry \citep{Evans2007}.

We performed point-spread function (PSF) photometry on the WFC3 imaging using DOLPHOT \citep{Dolphin2000, Dolphin2016}.
First, we downloaded the individual exposures (*flc.fits files) in all 5 filters processed with the default calwf3 pipeline from the Mikulski Archive for Space Telescopes (MAST).
To combine individual images into a deep mosaic required by DOLPHOT as a reference for alignment, we then used DrizzlePac v3.1.8 \citep{Gonzaga2012} to perform cosmic-ray rejection, alignment, and coaddition.
We used the F475W mosaic as the reference, as it is deeper than the NUV mosaics and lies in the middle of the wavelength range spanned by the observations.
Finally, we adopted the same settings optimized for photometry of bright O stars in nearby dwarf galaxies by \citet{Telford2021} in the DOLPHOT parameter file. 

DOLPHOT measures the brightness of stars in the individual images, aligned to the deep reference image, then combines those individual measurements to produce a catalog of magnitudes and uncertainties. 
We only included stars detected at signal-to-noise ratio (S/N) $>4$ in our photometric catalog.
We also used quality parameters in the DOLPHOT output to filter sources from our catalog for which the photometry is poorly recovered (FLAG parameter), impacted by nearby sources (CROWD parameter), or whose shapes deviate from the expectation for a stellar source (SHARP parameter).
Following the quality cuts developed by the Panchromatic Hubble Andromeda Treasury (PHAT) team to ensure high-quality data and remove spurious sources from the DOLPHOT output \citep{Williams2014}, we required FLAG $<4$ in all filters, CROWD $<1.3$ and SHARP$^2 < 0.15$ in NUV filters, and CROWD $<2.25$ and SHARP$^2 <0.2$ in the optical filters.
However, the sources in our final sample are all bright, and have higher photometric quality than the minimum requirements imposed on the full catalog at this stage.

We present the photometry for the 25 observed targets in our sample (selected as described in the following section) in \autoref{tab:phot} in \autoref{app:phot}.

\subsection{Optical Spectroscopy}\label{subsec:spec}

We selected candidate OB stars for optical spectroscopic follow-up based on their HST photometry, requiring stars to be brighter than {18.5 mag} in F275W and have F275W -- F336W colors bluer than $-0.2$ mag (see \autoref{fig:selection_cmd}). We excluded the two brightest stars that met these criteria, which are known B supergiants \citep{Evans2007} as we are primarily interested in the candidate O stars. This resulted in a sample of 44 candidate OB stars, many of which had not been classified spectroscopically before.

We obtained optical spectroscopy of our new and previously classified OB-star candidates in NGC 3109 using the Deep Imaging Multi-Object Spectrograph \citep[DEIMOS;][]{Faber2003} on the Keck II telescope on the nights of March 20 and 21, 2023 (NASA Keck program 2023A\_N048; PI: O.\ G.\ Telford). We used a single slitmask, which covers an area of 16\arcmin$\times$ 5\arcmin, convenient for observing NGC 3109, which has an apparent size of 19.1\arcmin$\times$ 3.7\arcmin. However, it was not possible to fit all of the candidate OB stars on one mask. Instead, we used the \texttt{dsimulator}\footnote{\url{https://www2.keck.hawaii.edu/inst/deimos/dsim.html}} software to design the DEIMOS slitmasks, prioritizing the candidates with bluer NUV colors and the three ULLYSES targets. The final masks included 25 of the 44 OB candidates.

We used the 1200B grating centered at 5000\,Å in order to capture the main diagnostic lines for O stars: the hydrogen Balmer series, \hei, and \heii\ transitions that fall between 4000--5000\,\AA\@. We used 0.75\arcsec\ slits with the GG400 order-blocking filter, resulting in a resolving power of $R\sim3750$ at 5000\,Å, or a FWHM of $\sim$1.3\,\AA\@. This is sufficient resolution to distinguish hydrogen Balmer absorption from nebular emission, a requirement for accurately estimating a star's surface gravity. The conditions were partially cloudy on both nights, with seeing ranging from 1--1.2\arcsec. We took 12 exposures of 20 minutes each -- five on the first night and seven on the second -- for a total of four hours of exposure time\footnote{The DEIMOS data presented in this paper can be downloaded from the Keck Observatory Archive: \url{https://koa.ipac.caltech.edu/}}. 

We reduced the spectra using \texttt{spec2d} \citep{Cooper2012,Newman2013}, a software pipeline written in IDL\@.  The pipeline produces wavelength-calibrated, sky-subtracted, one-dimensional spectra.  We used some modifications to the pipeline.  Foremost, \citet{delosReyes2020} improved the wavelength solution of DEIMOS 1200B spectra with a more liberal selection of arc lines.  Their improvement also allows the identification of arc lines by element to eliminate misidentification of arc lines.  For that reason, we took separate arc lamp exposures of Ne, Ar, Hg, and a simultaneous exposure of Kr and Xe, whereas the standard procedure is to turn on all lamps simultaneously.  The lines were identified exposure by exposure.  \citet{Kirby2015a} also improved the tracing of sky lines to increase the precision of the wavelength calibration.  Finally, \citet{Kirby2015b} changed the shape of the spectral extraction window to match the curvature from differential atmospheric refraction, thereby increasing the S/N of the extracted spectrum.  We used all of these updates.

Despite the improvements to the spectral reduction, the wavelength calibration was unreliable for some slits.  The primary reason for the unreliability was that few arc lines are available for the blue wavelengths accessed by the 1200B grating.  As a result, the polynomial fit from pixel position to wavelength can become poorly constrained at the edges of the spectrum, including the edges of the red or blue CCD near the chip gap, and even the middle of the spectrum (around 5000\,\AA) can sometimes have poor wavelength calibration.  Given the difficulty of deriving precise wavelength calibrations for the 1200B grating, we do not provide absolute radial velocities in this paper, but we note that the apparent velocities of the absorption lines we measure in Section~\ref{sec:fitting} are consistent within each spectrum and so we are not concerned about the impact of the wavelength calibration on the other derived parameters. 

We normalized the spectra by their stellar continuum, which we approximated with a cubic spline function. The continuum for the blue and red halves of the spectra were fit separately, with absorption and emission lines in the spectra masked prior to the spline fitting. There is often a flux discontinuity across the CCD chip gap in DEIMOS, which makes it prudent to fit the continua separately in each half of the spectrum.

\begin{figure*}[t]
\begin{center}
\includegraphics[width=1\linewidth,angle=0]{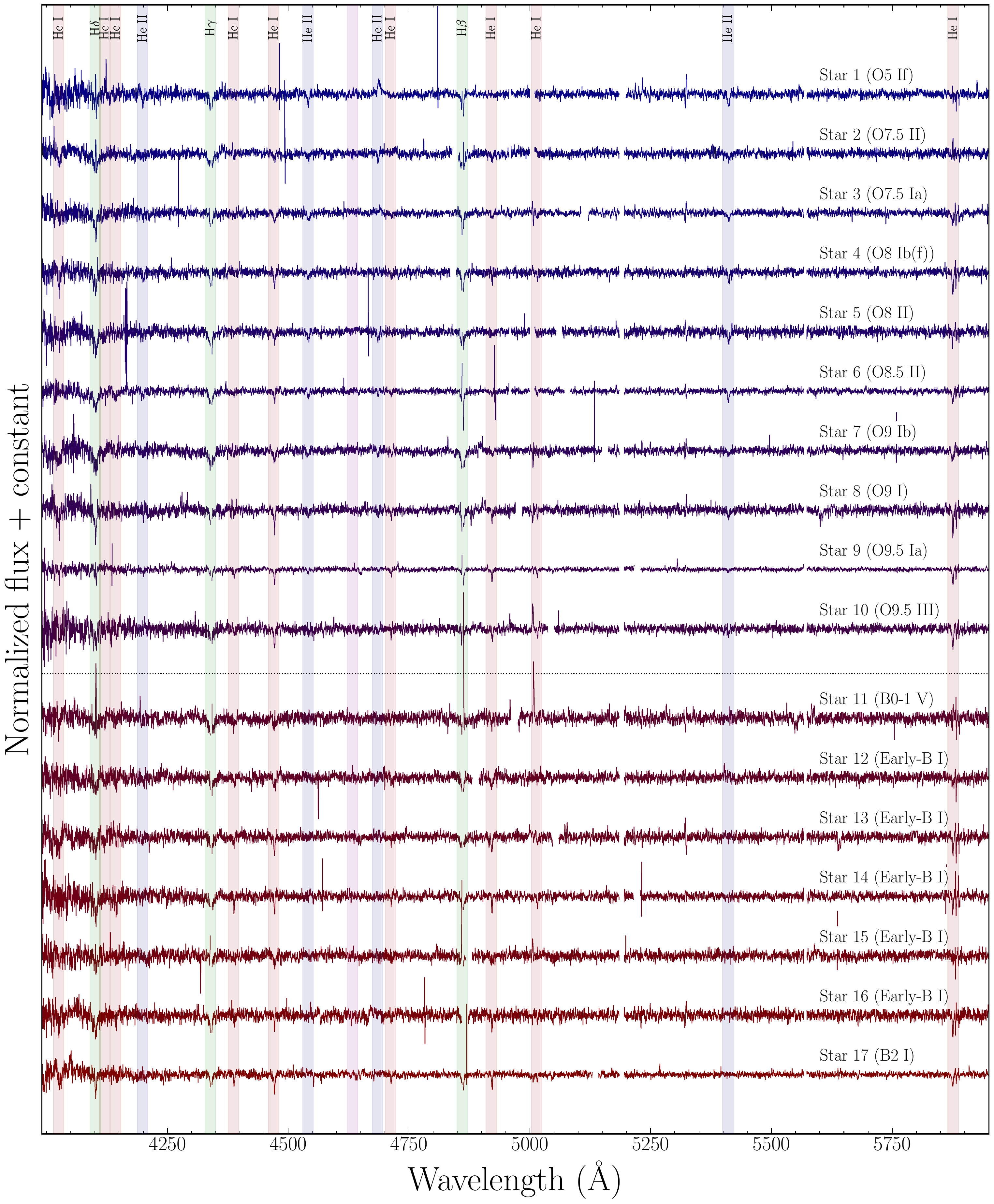}
\caption{Medium-resolution optical spectra for the 17 OB stars in our sample to which we assign spectral types. The stars are ordered and colored by spectral type, with the earliest-type stars on the top plotted in blue and the latest-type stars on the bottom shown in red. The dotted black horizontal line divides the O stars and B stars. Some of the key lines used in the classification and spectral fitting are highlighted. \hei\ transitions are shaded red, \heii\ transitions are shaded blue, and Balmer lines are shaded green.}\label{fig:3_allspec}
\end{center}
\end{figure*}

Ultimately, we obtained spectra of 25 OB-star candidates in NGC 3109. However, due to the poor conditions and faintness of the objects, we deemed eight of the spectra to be too noisy for a reliable analysis. These eight spectra (with an average SN of 12) had few, if any, helium lines detected with EW/$\sigma_\textrm{EW}>2$, making them impossible to classify. The remaining 17 spectra are shown in \autoref{fig:3_allspec}.

\begin{table*}[t]
    \caption{OB stars in NGC 3109}\label{tab:class}
    \centering
\begin{tabular}{cccccccc}
\hline\hline

     Star ID & R.A. & Dec. & Spectral Type & $\langle$S/N$\rangle$ & Previous ID & Previous Spectral Type & E(B--V)$_\textrm{fg}$\\
     (1) & (2) & (3) & (4) & (5) & (6) & (7) & (8)\\
\hline
1	 & 10:02:56.234	 & -26:08:58.22	 & O5 If	 & 22	 & EBU48	 & Late-O If	 & 0.067 	\\
2$^a$	 & 10:03:12.517	 & -26:08:51.66	 & O7.5 II	 & 20	 & --	 & --	 & 0.067 	\\
3	 & 10:03:02.515	 & -26:09:36.02	 & O7.5 Ia	 & 26	 & EBU33	 & O9 If	 & 0.067 	\\
4$^b$	 & 10:03:14.304	 & -26:09:17.89	 & O8 Ib(f)	 & 18	 & EBU34	 & O8 I(f)	 & 0.066 	\\
5$^a$	 & 10:03:01.338	 & -26:08:38.05	 & O8 II	 & 21	 & --	 & --	 & 0.067 	\\
6$^{a,b}$	 & 10:03:03.294	 & -26:09:21.33	 & O8.5 II	 & 31	 & EBU20	 & O8 I	 & 0.067 	\\
7	 & 10:02:59.739	 & -26:09:24.32	 & O9 Ib	 & 22	 & EBU49	 & O9 II	 & 0.067 	\\
8	 & 10:03:11.770	 & -26:10:18.94	 & O9 I	 & 18	 & EBU43	 & O9.5 I	 & 0.055 	\\
9$^b$	 & 10:02:54.718	 & -26:08:59.66	 & O9.5 Ia	 & 45	 & EBU07	 & B0-1 Ia	 & 0.067 	\\
10	 & 10:03:01.406	 & -26:08:26.61	 & O9.5 III	 & 18	 & EBU42	 & B0-2	 & 0.067 	\\
11$^c$	 & 10:03:12.366	 & -26:10:21.33	 & B0-1 V	 & 14	 & --	 & --	 & 0.055 	\\
12	 & 10:03:17.923	 & -26:10:53.88	 & Early-B I	 & 15	 & --	 & --	 & 0.044 	\\
13	 & 10:03:08.173	 & -26:10:19.13	 & Early-B I	 & 17	 & EBU60	 & B0-0.5	 & 0.059 	\\
14	 & 10:02:58.640	 & -26:09:57.98	 & Early-B I	 & 21	 & EBU44	 & Early-B Ib	 & 0.067 	\\
15$^c$	 & 10:03:16.630	 & -26:10:21.80	 & Early-B I	 & 15	 & EBU61	 & Early-B	 & 0.053 	\\
16	 & 10:03:14.114	 & -26:09:33.13	 & Early-B I	 & 14	 & --	 & --	 & 0.063 	\\
17	 & 10:02:59.926	 & -26:09:12.53	 & B2 I	 & 28	 & EBU22	 & B1 Ia	 & 0.067 	\\
\end{tabular}
	
    \tablecomments{The sample of 17 OB stars to which we assign spectral types. Column (1): the identification number we assign to each star, ordered by spectral type. Column (2): right ascension in hours, minutes, and seconds. Column (3): declination in degrees, arcminutes, and arcseconds. Column (4): spectral type as classified in Section~\ref{sec:spectralclass}. Column (5): average signal-to-noise ratio of the optical spectra. Column (6): identification number of each star from \citet{Evans2007}. Column (7): spectral type from \citet{Evans2007}. Column (8): foreground color excess from \citet{Schlegel1998} with updated calibrations in \cite{Schlafly2011} accessed using \texttt{dustmaps} \citep{Green2018}. $^{(a)}$Binary candidate. $^{(b)}$ULLYSES target. $^{(c)}$Stars with no reported stellar parameters.}
\end{table*}

\section{Spectral Classification} \label{sec:spectralclass}

The spectral classification of the OB stars in our sample is based primarily on the classification schemes presented in \citet{Mathys1988}, \citet{Sota2011}, \citet{Sota2014}, \citet{Martins2017}, and \citet{Evans2015}, which rely on the relative strengths of absorption lines (primarily of helium) in the blue optical wavelength range of the spectra. In addition to the line ratio criteria, we also visually compare our spectra to those from \citet{Walborn1990} and the GOSSS library \citep{Sota2011, Sota2014} for further confirmation. 

We initially separate the stars into O and B stars based on the presence and absence of \heii{}  absorption (detected with EW/$\sigma_\textrm{EW}>2$), respectively, as B-star atmospheres are not hot enough to produce significant \heii. {While very early B0 stars can exhibit weak \heii{} absorption, we confirm below using more detailed classifications that our initial separation is correct.} We find ten O stars and seven B stars. Below, we summarize our approach to assigning spectral types to the O and B stars. Additional details of the classification of each star are reported in \autoref{app:class} along with the measured EWs of the relevant lines in \autoref{tab:ews}. The assigned spectral types of the stars are presented in \autoref{tab:class}.

\subsection{O Stars}\label{subsec:classOstars}

To assign subtype classifications to the ten O stars in our sample, we follow \citet{Mathys1988}, \citet{Sota2011, Sota2014}, and \citet{Martins2018} and rely on the ratios of \hei{} to \heii{} absorption lines, which trace the helium ionization ratio and the temperature of the stellar atmosphere.  Our primary diagnostic is the \trans{he}{i}{4471}/\trans{he}{ii}{4542} ratio used by \citet{Mathys1988}, which we find to be the most robust; both of these lines are detected with EW/$\sigma_\textrm{EW}>2$ in all of the O-star spectra. We also consider the criteria presented in \citet{Martins2017} based on the ratios of \trans{he}{i}{4144}/\trans{he}{ii}{4200} and \trans{he}{i}{4388}/\trans{he}{ii}{4542}, finding them to be largely consistent with the \citet{Mathys1988} classifications for the late-type O stars. \citet{Martins2018} also presented classification criteria for late-type O stars based on the \trans{si}{iii}{4552}/\trans{he}{ii}{4542} ratio, but we do not consider this ratio as it is poorly understood whether classification criteria that are calibrated on Galactic standards and combine multiple elements can be consistently applied to lower-metallicity stars. We identify all except one of the O stars as late-type, exhibiting strong \hei{} absorption. We classify the O star with the earliest subtype as O5 and the remaining O stars as O7.5 -- O9.5.

Following \citet{Sota2011}, we assign LCs to the O stars based primarily on the strength of \trans{he}{ii}{4686} for the earlier O stars in our sample ($<$O8.5) and the ratio of \trans{he}{ii}{4686}/\trans{he}{i}{4713} for the later O stars (O9 -- O9.7). We classify all of the O stars in our sample as giants or supergiants. {We assign qualifiers of f and (f) to stars 1 and 4 respectively, to indicate possible \ion{N}{3} $\lambda$4634-40-42 emission and \trans{he}{ii}{4686} emission (for Star 1) or neutral \trans{he}{ii}{4686} (for Star 4). While the \ion{N}{3} $\lambda$4634-40-42 emission in these stars is weak, it is detected with S/N $>$ 2. Considering the metal-poor nature of these stars, and their notable \trans{he}{ii}{4686}, we feel these qualifiers are justified.} 

The spectral classes of the O stars are reported in \autoref{tab:class} along with previous classifications from \citet{Evans2007} for the stars which are included in their sample. Our classifications are generally similar to those from \citet{Evans2007}, with a few exceptions. We classify Star~3 as an earlier-type O star than \citet{Evans2007} and we classify Stars 9 and 10 as late-type O stars, while \citet{Evans2007} assigned Early-B types to both stars. These changes in spectral type are likely a result of our higher resolution spectroscopy, as we discuss in more detail in \autoref{app:class}.

\subsection{B Stars}\label{subsec:classBstars}
To assign spectral subtypes to the B stars, we follow the classification scheme of \citet{Evans2015}, which is designed to place B stars on a temperature sequence based on the relative strengths of silicon, magnesium and helium absorption lines. The \citet{Evans2015} criteria were calibrated to stars in the LMC. The LMC has a gas phase metallicity of $\sim0.5\,Z_\odot$, higher than NGC 3109, but more appropriate than classification criteria based on Galactic standards. We keep this in mind when classifying the B stars.

First, we assign each star a LC by comparing the width of the Balmer lines to standard spectra of dwarf and giant B stars from \citet{Walborn1990} and the GOSSS library \citep{Sota2011, Sota2014}. Giant stars are expected to have narrower absorption lines as compared to dwarf stars, which have higher surface gravity. We classify all but one of the B stars as giants, unsurprising given that B dwarfs would likely be too faint to observe. {While the width of the Balmer lines is dependent on temperature as well as surface gravity (and therefore varies with both LC and spectral type), we find the LCs of the B stars in our sample are largely unambiguous. In the stars classified as giants, the Balmer lines of the standard dwarf stars are all much broader than the lines in our spectra.} 

After assigning each star a LC, we determine its subtype classification based on the criteria presented in \citet{Evans2015}, {which is different for giants and dwarfs}. Given that these criteria are largely qualitative and rely on the relative strengths of metal lines, which are often weak for our metal-poor stars, our B-star subtypes are only approximate. The challenge of classifying the B-stars is compounded by the fact that these stars are the dimmest in our sample, and noisier on average than the O stars. For some of the low-SN spectra, we are only able to classify the stars as ``Early B." Nevertheless, we find that our assigned classifications largely agree with those previously presented in \citet{Evans2007}. {We also compare our B star classifications to the criteria from \citet{Castro2008} and find that they generally are consistent with the spectral types based on \citet{Evans2015}.}

\section{Stellar parameters from optical spectroscopy and photometry} \label{sec:fitting}

In this section, we fit stellar atmosphere models to the optical spectroscopy and HST photometry of the OB stars in our sample in order to estimate their fundamental stellar parameters. In Section~\ref{subsec:powr}, we describe the models and the precomputed grids of model spectroscopy and SEDs that we employ in our analysis. In Section~\ref{subsec:fitting_spec} we detail the procedure we follow to fit the optical spectroscopy to determine $T_\star$, $\log(g)$, and $v\sin i$. In Section~\ref{subsec:fitting_phot}, we compare the observed photometry for each star to the synthetic photometry predicted by the model that is best-fit to its spectroscopy to estimate the reddening towards each star, the stellar luminosity, radius, and mass. 

\subsection{PoWR Model Grid} \label{subsec:powr}

\begin{figure*}[t]
\begin{center}
\includegraphics[width=\linewidth,angle=0]{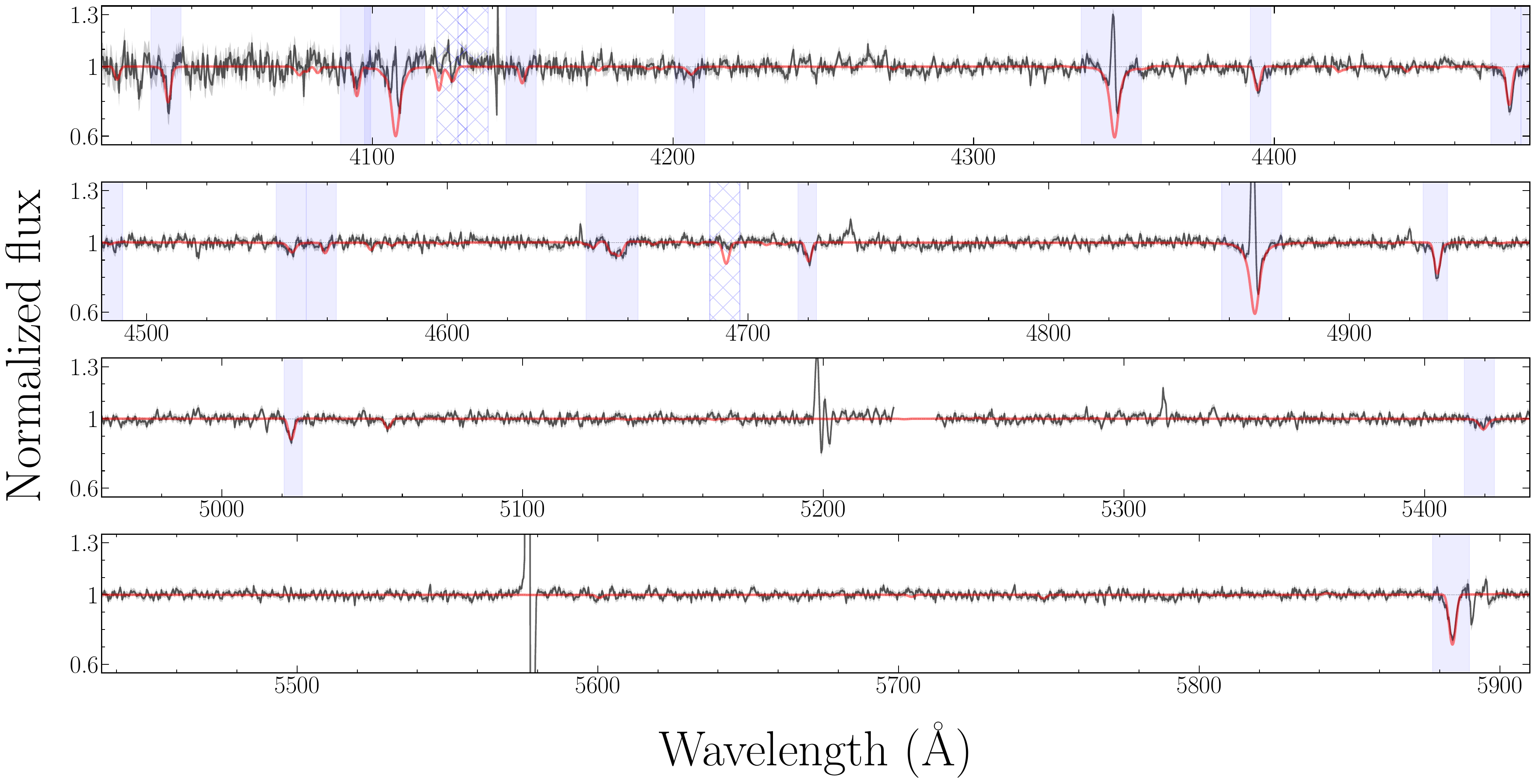}
\caption{The PoWR model atmosphere fit to the optical spectrum of Star 9, an O9.5 supergiant. The black line is the stellar spectrum, the gray shaded region is the one $\sigma$ uncertainty on the spectrum, and the red line is the PoWR model spectrum. The blue shaded regions represent lines that are included in the spectral fitting and the blue hatched regions are lines that are excluded from the fitting. A complete list of lines included in the model fitting is given in \autoref{app:allfits}.}
\end{center}
\end{figure*}

To estimate the fundamental stellar parameters $T_\star$, $\log(g)$, and $v\sin i$ from our observed optical spectroscopy, we rely on grids of model spectra generated by the Potsdam Wolf-Rayet code \citep[PoWR;][]{Grafener2002, Hamann2003, Sander2015}. PoWR is an advanced stellar atmosphere code which simultaneously solves the radiative transfer and statistical equilibrium equations without assuming local thermal equilibrium (i.e. non-LTE). 

While the PoWR code can be used to generate model spectra and SEDs for any choice of fundamental parameters $T_\star, L_\star, M_\star$ (or $\log (g)$ in place of $M_\star$), we instead rely on the publicly available precomputed grids of model atmospheres, which span a range of $\log(g)$ and $T_\star$ \citep{Hainich2019}\footnote{\url{https://www.astro.physik.uni-potsdam.de/~wrh/PoWR/}}. We note that the PoWR models are computed using $T_\star$, the temperature corresponding to the stellar radius $R_\star$, or more specifically, the radius at which the Rosseland optical depth $(\tau_\text{Ross})$ is equal to 20. The effective temperature $T_\text{eff}$, which corresponds to $\tau_\text{Ross}=2/3$, is often used in the literature and is reported for each PoWR model. However, given that the model grid is computed using $T_\star$ and the typical difference between the two temperatures is $\sim100$ K (a small fraction of our typical uncertainty), we report and use values of $T_\star$ in this work.

The PoWR team provides several precomputed grids of model spectra and SEDs for OB stars: one at Galactic metallicity, one at LMC metallicity, and four at SMC metallicity -- each with a different prescription for mass-loss rates\footnote{The SMC models' metallicity is reported as 1/7 $Z_\odot$, but the oxygen mass fraction of 0.0011 used in their computation corresponds to a metallicity of 19\% $Z_\odot$ with respect to the solar oxygen abundance reported in \citet{Asplund2009}.}. For the SMC-OB-VD3 grid, $\dot{M_\star}$ is calculated for each model following \citet{Vink2001}, but reducing the calculated value by a factor of three to better match observational constraints \citep{Ramachandran2019, Rickard2022}. The mass-loss rates for the remaining SMC grids are determined based on a fixed value of the wind strength parameter $Q$, defined as $Q= \dot{M_\star}(R_\star v_\infty)^{-3/2}$. For the PoWR grids SMC-OB-I (moderate), SMC-OB-II (high), and SMC-OB-III (low) $\log(Q)$ is fixed as $-13, -12$, and $-14$ respectively\footnote{{In January 2025, additional PoWR grids were made public, including an updated version of the SMC-OB-VD3 grid (covering a larger set of $T_\star$-$\log(g)$ values) and a new grid with VD3 mass loss at 3\%\,$Z_\odot$. We have not included these new grids in our analysis, in part to maintain consistency with the high-mass-loss grid, which had not been updated at the time of writing.}}. 

For our analysis, we consider two grids with SMC metallicity, which is the closest of the available metallicities to that of NGC 3109. We test the sensitivity of the fitted parameters to the chosen metallicity in Section~\ref{subsec:fitting_spec} and \autoref{app:tlusty}.

We use the SMC-OB-VD3 grid as our default model grid, but in some cases prefer the models of the high mass-loss rate grid (SMC-OB-II) for stars with evidence of elevated mass loss. The details of this model choice are discussed further in the following section.

The SMC-OB-VD3 and SMC-OB-II grids include 227 and 194 different models respectively, with $T_\star$ ranging from 15 kK to 50 kK and $\log(g/\rm{cm \ s}^{-2})$ ranging from 2.0 to 4.4. The SMC-OB-VD3 grid includes models with high surface gravity and low temperature that are not included in the SMC-OB-II grid, as these models are unstable with such high mass loss rates. The models in each grid are spaced by 1\,kK in $T_\star$ and 0.2\,dex in $\log(g)$. 

For each model spectrum, we apply a rotational broadening kernel with values of $v\sin i$ ranging from 0 to 440 km s$^{-1}$ spaced by 40 km s$^{-1}$, with the spacing chosen to be half the spectral resolution. We also convolve the spectra with a gaussian kernel with FWHM of 1.35 \AA\ to match the resolution of the instrument. 

\subsection{$T_\star$ and $\log$(g) from Optical Spectra} \label{subsec:fitting_spec}
\begin{figure*}[t]
\begin{center}
\includegraphics[width=\linewidth,angle=0]{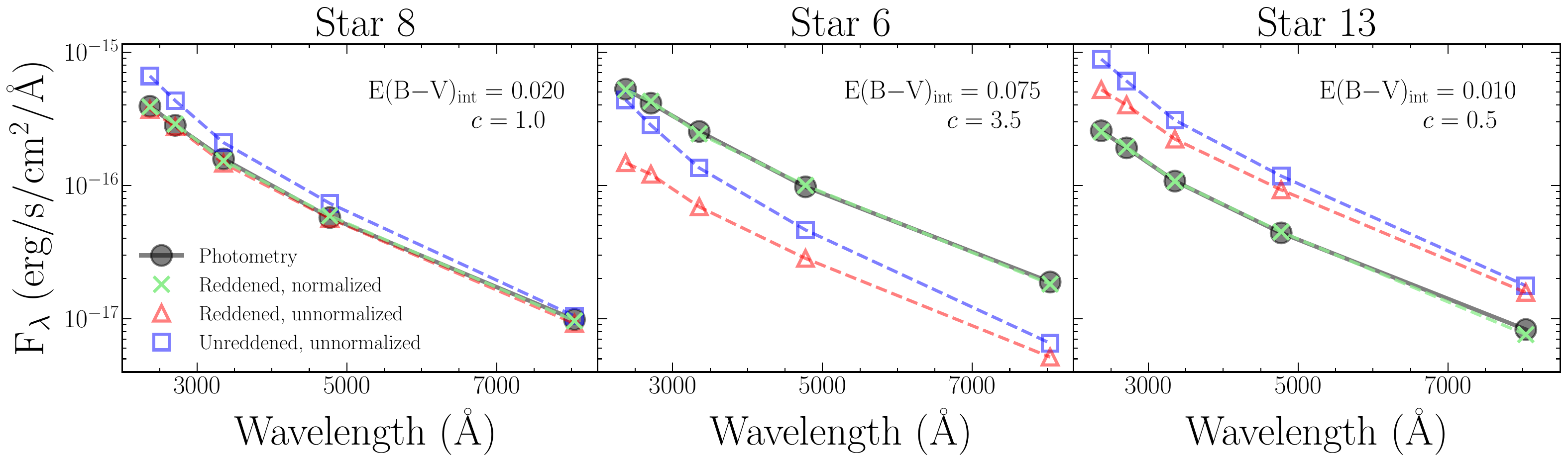}
\caption{Three examples of the procedure for estimating the color excess due to dust internal to NGC 3109 [$E(B-V)_\text{int}$] and the stellar luminosity (L$_\star$) for Stars 8, 6, and 13 (from left to right). The black circles show the HST photometry for each star. The blue squares show the synthetic HST photometry measured from the unreddened, unnormalized SED predicted for the PoWR model which is best fit to each star's optical spectroscopy. The red line shows the synthetic photometry from the reddened but unnormalized SED and the green line the synthetic photometry from the reddened and normalized SED. In the left most panel, the green and red lines are perfectly overlapping ($c=1$), implying that the predicted SED matches the observed photometry perfectly with only a small amount of reddening applied. The photometry of the star in the middle panel is significantly brighter than what is predicted by its best fit stellar parameters. This star is identified in Section~\ref{subsec:stellarparams} as a binary candidate. The photometry of the rightmost star is slightly fainter than what is predicted by its best-fit parameters.}
\end{center}
\end{figure*}

To determine the temperature, surface gravity, and projected rotational velocity for each of our stars, we fit the model grids to the optical spectra using a $\chi^2$ minimization. We evaluate each model based only on its fit to stellar absorption lines in the spectra. Our default approach is to fit all \hei{}, \heii{}, and Balmer lines. We also include a number of Mg and Si transitions, unless the regions of the spectra containing those transitions are extremely noisy. In a small number of cases, we see evidence of C or N absorption or emission and include those transitions as well. We exclude the \trans{he}{ii}{4686} line from the fit as this line is strongly affected by stellar winds and we do not expect the models to fit the observed profile. We visually inspect each spectrum and exclude additional lines if they are dominated by noise, contaminated by emission, or otherwise unreliable. A full list of lines included in the fitting of each star is provided in \autoref{app:allfits}.

After determining the appropriate set of lines for each star, we calculate the reduced $\chi^2$, $\chi^2_\nu=\chi^2/\nu$, for each combination of $T_\star,\ \log$(g), and $v\sin i$. The reported $\chi^2_{\nu}$ for each model is calculated as
\begin{equation}
    \chi^2_{\nu} = \frac{1}{\nu}\sum_i\left[\frac{s(\lambda_i)-m(\lambda_i)}{\sigma(\lambda_i)}\right]^2  
\end{equation}
where $s(\lambda_i)$ is the continuum-normalized flux from the observed optical spectrum at wavelength $\lambda_i$, $m(\lambda_i)$ is the continuum-normalized flux from the model spectrum at wavelength $\lambda_i$, $\sigma(\lambda_i)$ is the error on the observed flux at wavelength $\lambda_i$, and $\nu$ is the number of degrees of freedom (i.e., the number of free parameters subtracted from the number of data points included in the fit). The model spectrum is redshifted to the radial velocity $v_r$ that minimizes $\chi^2_{\nu}$ for that model.

The preferred model is chosen as that with the overall lowest value of $\chi^2_{\nu}$. {To estimate the confidence intervals for our reported values of $T_\star$, $\log$(g) and $v\sin i$, we determine the probability that the $\chi^2$ of a given model deviates from the minimum $\chi^2$ solely due to random noise by calculating $Q=1-\Gamma(\nu/2, \chi^2/\chi^2_{\nu,\textrm{ min}}/2)$ with $\Gamma$ the lower incomplete gamma function. The $\chi^2$ values are normalized by the minimum $\chi^2_\nu$ to minimize possible bias introduced by incorrectly estimated errors on the flux. We define good-fitting models as those with $Q\geq 0.32$ -- i.e. those with a greater than 32\% probability of deviating from the best-fit model model due to random fluctuations. The confidence interval for each parameter is then taken to span the range of values covered by the good-fitting models. In the event that the best-fit value is at either extreme of the confidence interval such that the reported uncertainty would be zero, we impose a minimum uncertainty of one grid spacing: 1\,kK for $T_\star$, 0.2 dex for $\log g$, and 40 km s$^{-1}$ for $v\sin i$. In some cases, the best-fit model is at the edge of the model grid and no model exists which is one grid spacing adjacent to the best-fit model. In these cases, we report an upper or lower uncertainty of zero to indicate that the model is at the grid boundary.}

We fit all of the optical spectra twice, once with the SMC-OB-VD3 grid and once with the higher mass-loss SMC-OB-II grid. In most cases, we report the stellar parameters of the best-fit model from the SMC-OB-VD3 grid. However, for two of the O stars (Stars 1 and 3), we observe \heii\ 4686 emission, a sign of strong winds and so report the values from the high mass loss grid for these two stars. The optimal $\chi^2_\nu$ is similar for the fits to the two models for these stars. 

We report the best-fit parameters and our derived confidence intervals for the O and B stars in \autoref{tab:stellarparams}. We present the fitted parameters for all stars from both grids in \autoref{app:powr}. For the majority of the stars, there is little or no difference between the best-fit parameters from the two grids. In some of the stars, including those we described above for which we prefer the higher mass-loss models, the parameters are slightly different, with the high mass-loss grids preferring models with higher temperatures and surface gravity. 

To test the sensitivity of our fitted parameters to the choice of model metallicity, we repeat the fitting procedure for a different set of model grids generated by the TLUSTY stellar atmosphere code. TLUSTY grids are available for O stars \citep{Lanz2003} and B stars \citep{Lanz2007} with metallicities ranging from 0 $Z_\odot$ to 2 $Z_\odot$. This wide range of metallicities allows us to test the impact of using higher or lower metallicity models to fit our spectra, but we ultimately prefer the fits from the PoWR models as the TLUSTY code does not account for the effects of stellar winds. We find that the TLUSTY SMC-metallicity grids yield similar values to the PoWR models and that there is little change when using grids with 0.1 $Z_\odot$. For the most part, there is also little difference when we use the LMC metallicity grids, except in a small number of cases where the LMC metallicity grids clearly overpredict the strength of metal lines in the spectra. We conclude that our use of SMC-metallicity PoWR grids for our analysis is well justified even though the metallicities of the stars in our sample may not be equal to that of the SMC. Details of the TLUSTY fitting are provided in \autoref{app:tlusty}.

\subsection{$L_\star$, $R_\star$, $M_{\star, \text{spec}}$ from Photometry}\label{subsec:fitting_phot}

\begin{table*}[t]
    \caption{Best-fit parameters from PoWR model fits to optical spectroscopy and NUV photometry.}\label{tab:stellarparams}
    \centering
\begin{tabular}{ccccccccccc}
\hline\hline
Star ID & T$_\star$ & $\log g$      & $v\sin i$       & $E(B-V)_\text{int}$ & c & $\log(L_\star)$ &    R$_\star$ & M$_{\star,\text{spec}}$ &  M$_{\text{evo}}$ &  M$_{0}$\\   
      & (kK)      & (cm s$^{-2}$)  & (km s$^{-1}$)                     &                     &   &  $L_\odot$          &    R$_\odot$               & M$_\odot$  & M$_\odot$ & M$_\odot$ \\  
(1) & (2) & (3) & (4) & (5) & (6) & (7) & (8) & (9) & (10) & (11)\\
\hline 
1$^a$ & 	45$_{-3}^{+1}$ & 	4.0$_{-0.2}^{+0.2}$ & 	80$_{-80}^{+80}$ & 	0.015$_{-0.010}^{+0.005}$ & 	1.4$_{-0.6}^{+1.8}$ & 	5.84$_{-0.10}^{+0.03}$ & 	14$_{-1}^{+1}$ & 	68$_{-23}^{+44}$ & 	57$_{-9}^{+4}$ & 	59$_{-9}^{+4}$ \\
2 & 	34$_{-2}^{+2}$ & 	4.0$_{-0.2}^{+0.2}$ & 	280$_{-40}^{+120}$ & 	0.000$_{-0.000}^{+0.005}$ & 	4.0$_{-2.2}^{+4.7}$ & 	5.40$_{-0.06}^{+0.07}$ & 	14$_{-1}^{+1}$ & 	75$_{-31}^{+54}$ & 	30$_{-2}^{+3}$ & 	30$_{-3}^{+4}$ \\
3$^a$ & 	32$_{-1}^{+2}$ & 	3.4$_{-0.2}^{+0.2}$ & 	160$_{-40}^{+40}$ & 	0.030$_{-0.015}^{+0.010}$ & 	1.2$_{-0.7}^{+2.1}$ & 	5.64$_{-0.06}^{+0.08}$ & 	21$_{-1}^{+1}$ & 	42$_{-16}^{+31}$ & 	38$_{-3}^{+5}$ & 	39$_{-3}^{+5}$ \\
4 & 	32$_{-1}^{+1}$ & 	3.4$_{-0.2}^{+0.2}$ & 	80$_{-40}^{+40}$ & 	0.000$_{-0.000}^{+0.010}$ & 	1.2$_{-0.7}^{+1.7}$ & 	5.56$_{-0.03}^{+0.05}$ & 	20$_{-1}^{+1}$ & 	35$_{-11}^{+26}$ & 	35$_{-2}^{+2}$ & 	36$_{-2}^{+3}$ \\
5 & 	35$_{-2}^{+1}$ & 	3.8$_{-0.2}^{+0.2}$ & 	200$_{-80}^{+40}$ & 	0.000$_{-0.000}^{+0.000}$ & 	2.3$_{-1.3}^{+2.7}$ & 	5.52$_{-0.06}^{+0.06}$ & 	16$_{-1}^{+1}$ & 	56$_{-21}^{+37}$ & 	34$_{-3}^{+3}$ & 	35$_{-3}^{+3}$ \\
6 & 	33$_{-1}^{+2}$ & 	3.6$_{-0.2}^{+0.2}$ & 	120$_{-40}^{+40}$ & 	0.075$_{-0.010}^{+0.010}$ & 	3.5$_{-2.3}^{+4.4}$ & 	5.80$_{-0.05}^{+0.08}$ & 	24$_{-1}^{+2}$ & 	85$_{-33}^{+64}$ & 	47$_{-3}^{+6}$ & 	49$_{-4}^{+7}$ \\
7 & 	34$_{-2}^{+1}$ & 	4.2$_{-0.2}^{+0.0}$ & 	240$_{-40}^{+40}$ & 	0.015$_{-0.010}^{+0.000}$ & 	9.2$_{-4.9}^{+1.1}$ & 	5.48$_{-0.08}^{+0.03}$ & 	16$_{-1}^{+1}$ & 	146$_{-61}^{+7}$ & 	33$_{-4}^{+2}$ & 	33$_{-4}^{+2}$ \\
8 & 	31$_{-2}^{+1}$ & 	3.4$_{-0.2}^{+0.2}$ & 	40$_{-40}^{+40}$ & 	0.020$_{-0.015}^{+0.005}$ & 	1.0$_{-0.7}^{+1.6}$ & 	5.41$_{-0.09}^{+0.04}$ & 	18$_{-1}^{+1}$ & 	28$_{-11}^{+21}$ & 	29$_{-3}^{+2}$ & 	30$_{-3}^{+2}$ \\
9 & 	27$_{-1}^{+1}$ & 	3.0$_{-0.0}^{+0.2}$ & 	80$_{-40}^{+40}$ & 	0.005$_{-0.005}^{+0.010}$ & 	1.1$_{-0.3}^{+2.3}$ & 	5.74$_{-0.05}^{+0.05}$ & 	34$_{-2}^{+2}$ & 	42$_{-4}^{+30}$ & 	41$_{-3}^{+3}$ & 	43$_{-3}^{+4}$ \\
10 & 	28$_{-2}^{+1}$ & 	3.2$_{-0.2}^{+0.2}$ & 	80$_{-40}^{+40}$ & 	0.005$_{-0.005}^{+0.010}$ & 	0.9$_{-0.6}^{+1.5}$ & 	5.36$_{-0.07}^{+0.06}$ & 	20$_{-1}^{+2}$ & 	24$_{-9}^{+18}$ & 	27$_{-4}^{+2}$ & 	28$_{-4}^{+2}$ \\
12 & 	20$_{-3}^{+4}$ & 	2.6$_{-0.2}^{+0.2}$ & 	200$_{-40}^{+40}$ & 	0.000$_{-0.000}^{+0.035}$ & 	0.3$_{-0.2}^{+0.5}$ & 	4.80$_{-0.08}^{+0.21}$ & 	21$_{-3}^{+6}$ & 	6$_{-3}^{+6}$ & 	15$_{-2}^{+4}$ & 	15$_{-2}^{+4}$ \\
13 & 	25$_{-1}^{+1}$ & 	3.0$_{-0.2}^{+0.2}$ & 	80$_{-40}^{+40}$ & 	0.010$_{-0.010}^{+0.010}$ & 	0.5$_{-0.4}^{+0.8}$ & 	5.08$_{-0.07}^{+0.06}$ & 	19$_{-1}^{+1}$ & 	13$_{-5}^{+9}$ & 	20$_{-2}^{+1}$ & 	20$_{-2}^{+1}$ \\
14 & 	26$_{-9}^{+1}$ & 	3.2$_{-0.8}^{+0.2}$ & 	80$_{-40}^{+40}$ & 	0.100$_{-0.100}^{+0.010}$ & 	1.5$_{-1.1}^{+2.1}$ & 	5.35$_{-0.50}^{+0.06}$ & 	23$_{-1}^{+8}$ & 	31$_{-23}^{+21}$ & 	26$_{-11}^{+3}$ & 	26$_{-11}^{+3}$ \\
16 & 	23$_{-4}^{+3}$ & 	3.0$_{-0.4}^{+0.2}$ & 	80$_{-40}^{+40}$ & 	0.010$_{-0.010}^{+0.040}$ & 	0.8$_{-0.6}^{+1.3}$ & 	5.05$_{-0.16}^{+0.18}$ & 	21$_{-1}^{+5}$ & 	16$_{-7}^{+12}$ & 	19$_{-3}^{+4}$ & 	19$_{-3}^{+4}$ \\
17 & 	18$_{-1}^{+3}$ & 	2.4$_{-0.0}^{+0.2}$ & 	80$_{-40}^{+40}$ & 	0.000$_{-0.000}^{+0.000}$ & 	0.5$_{-0.2}^{+1.2}$ & 	5.14$_{-0.02}^{+0.10}$ & 	38$_{-7}^{+5}$ & 	13$_{-2}^{+14}$ & 	20$_{-1}^{+3}$ & 	21$_{-1}^{+3}$
\end{tabular}
    \tablecomments{{The upper and lower limits reported in the table represent the 68\% confidence interval of the derived parameters as described in Section~\ref{sec:fitting}. Uncertainties of zero indicate that the best-fit model lies at the boundary of PoWR grids and no adjacent model exists.} Column (1): the identification number assigned to each star. Column (2): stellar temperature. Column (3): log surface gravity. Column (4): projected rotation speed. Column (5): color excess internal to NGC 3109 along the line of sight to each star. Column (6): normalization factor derived in fitting observed photometry to predicted SED. Column (7): stellar luminosity. Column (8): stellar radius. Column (9): spectroscopic stellar mass. Column (10): evolutionary mass. Column (11): initial evolutionary mass. $^{(a)}$ Parameters reported from high-mass-loss grid OB-SMC-II.}
\end{table*}

\begin{figure*}[t]
\begin{center}
\includegraphics[width=\linewidth,angle=0]{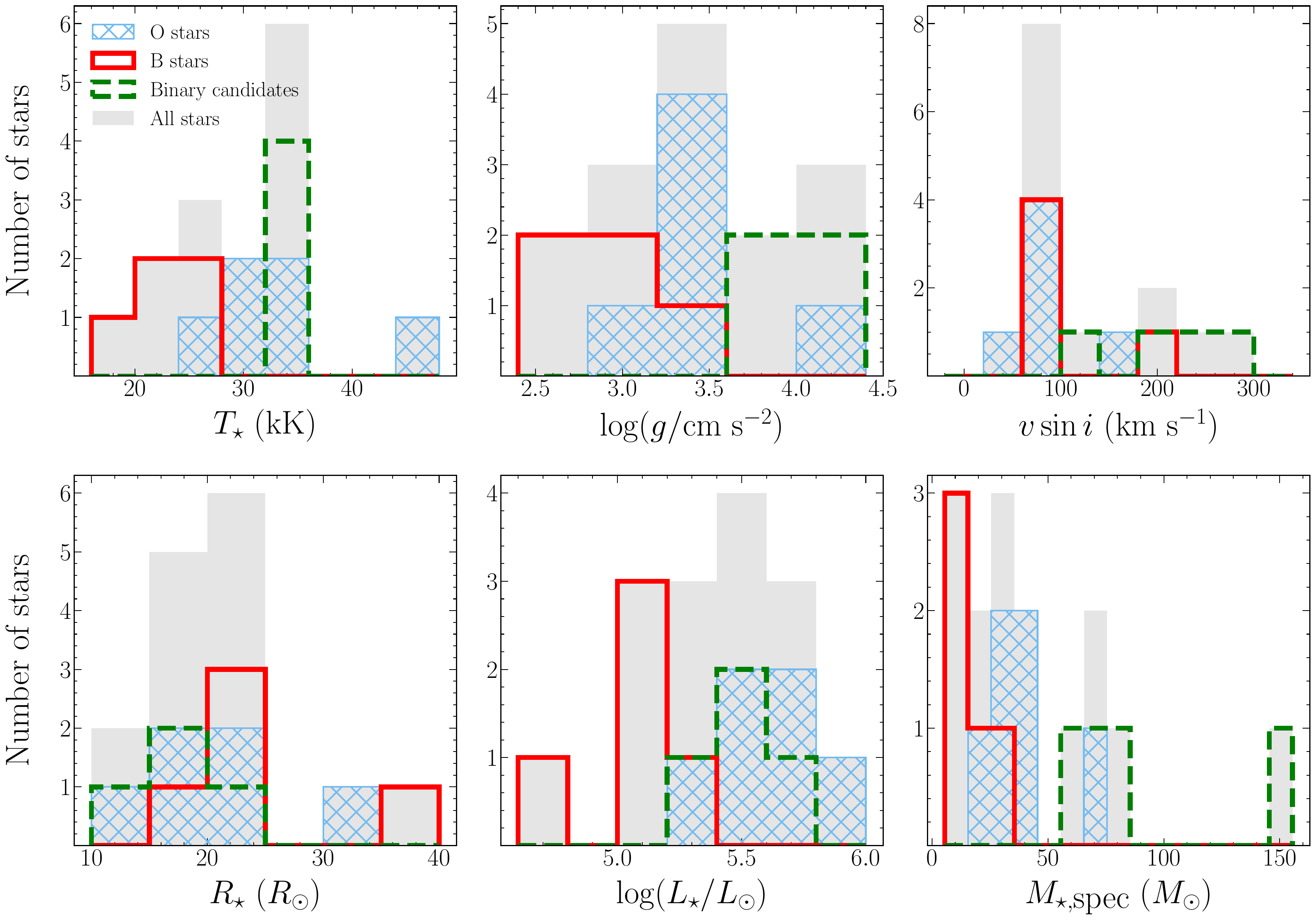}
\caption{Histograms showing the distributions of the stellar parameters measured by fitting PoWR model spectra and SEDs to our observed optical spectroscopy and optical and NUV photometry. The green dashed histograms show the distributions for the four binary candidates identified in Section~\ref{subsec:binary_sel}, the blue hatched histograms show the distributions for the remaining six O stars, the red histograms show the distributions for the five B stars, and the filled gray histograms show the distributions for all 15 stars together. The O stars are generally hotter, more massive, and more luminous than the B stars. The binary candidates have high surface gravities, high luminosities, high rotational speeds, and high spectroscopic stellar masses. \label{fig:paramhist}}
\end{center}
\end{figure*} 

\begin{figure*}[t]
\begin{center}
\includegraphics[width=\linewidth,angle=0]{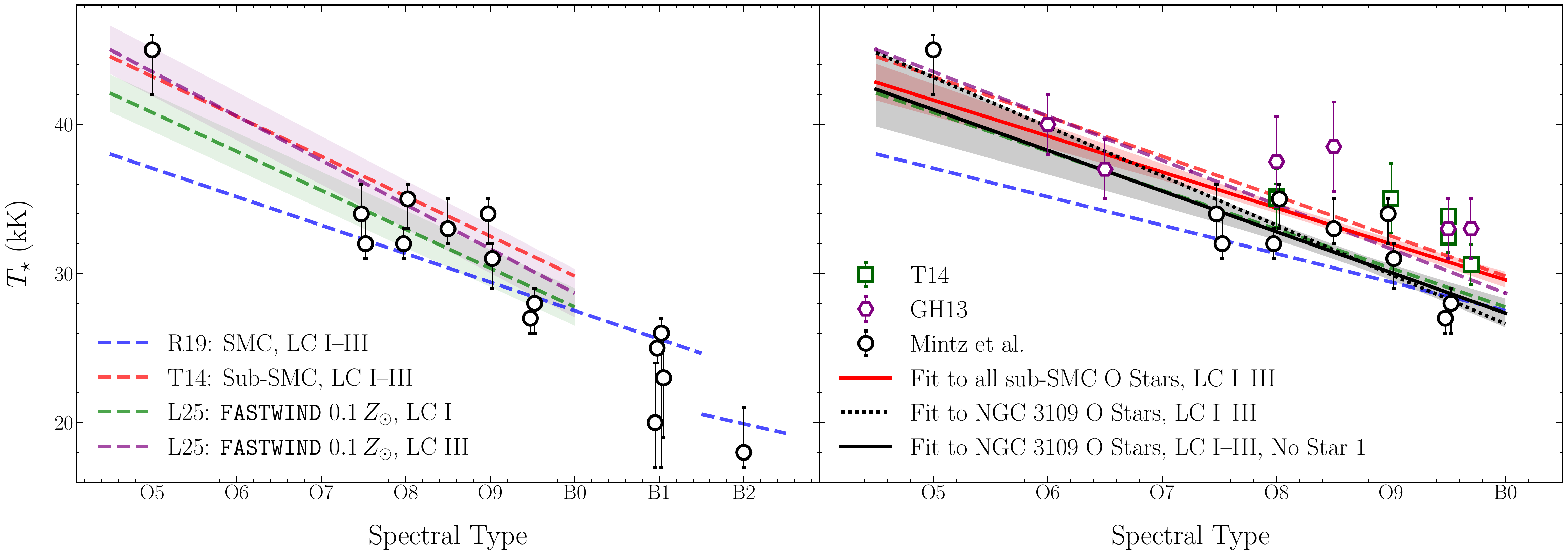}
\caption{(Left) The stellar temperatures $T_\star$ as a function of spectral type for the 15 OB giants, bright giants, and supergiants in our sample are shown in black. Stars with identical spectral types are shifted slightly for legibility and stars with type ``Early B" are placed at B1. Previously published relations are shown for SMC stars with LC I--III \citep{Ramachandran2019} and sub-SMC stars with LC I--III \citep{Tramper2014}. {Relations derived from model grids in \citet{Lorenzo2025} are shown in green for supergiants and in purple for giants.} (Right) The same as the left panel, but including additional sub-SMC O stars with temperatures and spectral types from \citet{Tramper2014} as green squares and from \citet{Garcia2013} as purple hexagons. In addition to the relations shown in the left panel, we also show the line fitted to the NGC 3109 stars (excluding Star 1) presented in this paper in black with one $\sigma$ confidence intervals shown as a gray shaded region. {We show the fit performed with Star 1 as a dotted black line.} The solid red line and red shaded region show the relation and confidence intervals fit to the full sub-SMC metallicity sample. {The late-type O giants in our sample are cooler than predicted by previous sub-SMC giant relations, possibly because they are mostly supergiants (LC I) whereas the O stars in \citet{Garcia2013} are largely giants (LC III). The relation fitted to the NGC 3109 O stars presented in this paper excluding Star 1 is nearly perfectly consistent with the model-derived relation for supergiants from \citet{Lorenzo2025}}. \label{fig:tspec}}
\end{center}
\end{figure*}

Once we have obtained estimates of $T_\star$ and $\log$(g) from the optical spectroscopy, we can combine these results with the HST photometry to estimate additional physical parameters of each star: stellar luminosity ($L_\star$), stellar radius ($R_*$), and spectroscopic stellar mass ($M_{\star, \textrm{spec}}$).

We first fit for the color excess towards each star $E(B-V)$ and the stellar luminosity $L_\star$ by comparing the HST photometry of each star with the spectral energy distribution (SED) predicted by the PoWR model that is best fit to its optical spectroscopy. The predicted SEDs can differ from the observed photometry for a number of reasons. The shape of the SED is affected by reddening due to dust along the line of sight. The luminosity could be higher than predicted because of unresolved binarity. The lack of low-metallicity constraints in the stellar evolution models used to set the stellar radius and luminosity in the PoWR model grids \citep{Brott2011} could also result in disagreement between predicted luminosities and observed values. 

For each star, we first obtain the SED of the model which provides the best fit to the optical spectroscopy as described in Section~\ref{subsec:fitting_spec}. We dim the SED appropriately to account for the 1.34 Mpc distance to NGC 3109. We then apply reddening, first using the dust extinction law from \citet{Gordon2023} with $R_V=3.1$ to account for foreground galactic extinction $E(B-V)_\text{fg}$, for which we use the dust maps of \citet{Schlegel1998} with updated calibrations  from \citet{Schlafly2011} accessed using \texttt{dustmaps} \citep{Green2018}. To estimate the internal reddening caused by dust within NGC 3109, we apply additional extinction using values for the color excess $E(B-V)_\text{int}$ ranging from 0 to 0.15 in increments of 0.005, resulting in 30 SEDs. We integrate each of the reddened SEDs through the appropriate photometric filters to produce synthetic HST photometry\footnote{Transmission curves for HST UVIS are taken from \url{https://www.stsci.edu/hst/instrumentation/wfc3/performance/throughputs} and synthetic photometry is generated using \texttt{sedpy} \citep{Johnson2021}.}. 

To determine the optimal value of $E(B-V)_\text{int}$ and L$_\star$, we again perform a $\chi^2$ minimization, fitting the reddened synthetic photometry to the observations. We permit each set of synthetic photometry to shift vertically to fit for the stellar luminosity -- or more specifically, to fit for the factor by which the true luminosity is offset from the luminosity adopted in the PoWR grid. The value of $\chi^2_\nu$ for a given model is calculated as
\begin{equation}
    \chi^2_\nu = \frac{1}{\nu}\sum_{i}\left(\frac{c\cdot\hat{F}_{i,E(B-V)_\text{int}}-F_i}{\sigma_i}\right)^2\label{eq:chi2_ebv}
\end{equation}
where $\hat{F}_{i,E(B-V)_\text{int}}$ is the synthetic flux through filter $i$ with an internal reddening of $E(B-V)_\text{int}$, c is the factor that allows for flexibility in the model SED normalization, $F_i$ is the measured flux through filter $i$, and $\sigma_i$ the error on the flux measurement through that filter. We take the best-fit value of $c$ and $E(B-V)_\text{int}$ for each star to be the values which produce the lowest $\chi^2_\nu$ for the SED of the model which is best fit to the spectroscopy. The luminosity of each star is then taken to be $L_\star = c\cdot L_{\star, \text{PoWR}}$ where $L_{\star, \text{PoWR}}$ is the stellar luminosity used in constructing the PoWR model SEDs. Similar to our approach in Section~\ref{subsec:fitting_spec}, we derive the uncertainty on the value of $E(B-V)_\text{int}$ and L$_\star$ by repeating the procedure described above for each of the models which satisfy {$Q>0.32$ and taking the range of values derived from these models to be the 68\% confidence interval.}

Using the best-fit $L_\star$ and $T_\star$ values, we then derive the stellar radius $R_\star$ using the Stefan-Boltzmann law:
\begin{equation}
    R_\star = \sqrt\frac{L_\star}{4 \pi \sigma_{SB}T_\star^4}
\end{equation}
where $\sigma_{SB}$ is the Stefan-Boltzmann constant.

Finally, we calculate the spectroscopic stellar mass, M$_{\star, \text{spec}}$, as
\begin{equation}
    M_{\star, \text{spec}} = \frac{g \cdot R_\star^2}{G}
\end{equation}
where g is the surface gravity of the best-fit model and G is the gravitational constant. The derived values of $E(B-V)_\textrm{int}$, $L_\star$, $R_\star$, and $M_{\star, \text{spec}}$ are reported in \autoref{tab:stellarparams}.

\subsection{Binary candidates}\label{subsec:binary_sel}
Four of the O stars in our sample have photometry that is significantly brighter than the SEDs predicted by the PoWR models which are best fit to their optical spectroscopy ($c>2$). It is possible that this discrepancy is a result of unresolved binarity (or multiplicity) in what we fit as single stars. We therefore flag these four stars (Stars 2, 5, 6, and 7) as candidate binary stars and discuss additional evidence for binarity in Section~\ref{subsubsec:binary}. Definitive classification of the binarity of these systems requires higher resolution spectroscopy or repeat observations to detect variability in radial velocity or photometry.

\section{Results} \label{sec:res}

\subsection{Stellar Parameters}\label{subsec:stellarparams}
Of the 17 OB stars that we classify in Section~\ref{sec:spectralclass}, we successfully measure stellar parameters for all but two of the B stars. We do not report measured parameters for Stars 11 and 15, nor do we include these two stars in subsequent analysis, because both of these B stars have only a small number of detectable lines and also have strong nebular emission in their Balmer lines, making it difficult to constrain their best-fit parameters via Balmer absorption. For the rest of the sample, we find that the parameters are reasonably well constrained. {For example, the median width of the reported confidence interval for T$_\star$ is 3\,kK - analogous to $\sigma\approx$ 1.5\,kK, which is comparable to typical reported temperature uncertainties for similar quality data \citep[eg.,][]{Tramper2014, Gull2022}.} The same is true for the rest of the reported parameters. 

We present histograms of our measured stellar parameters for the sample of 15 stars in \autoref{fig:paramhist}, with the gray filled histogram showing the distributions for the full sample. We include histograms for the O and B stars separately (shown as blue hatched and red unfilled histograms) and for the binary candidates (green dashed, unfilled histograms). As expected, we find that the O stars generally have higher temperatures, luminosities, and masses than the B stars. We also find that the majority of our stars (excluding the binary candidates which we discuss in detail below) have relatively low values of $\log(g)$ as compared to larger studies of SMC-metallicity OB stars \citep{Ramachandran2018b}. This concentration of low surface gravity agrees with our classification of all of these stars as giants or supergiants. 

The average $v\sin i$ for our sample is $\approx130$\,km\,s$^{-1}$, similar to the average projected rotational speed of $\approx150$\,km\,s$^{-1}$ reported in \citet{Ramachandran2019} for a sample of hundreds of OB stars in the SMC. \citet{Ramachandran2019} see evidence of a bimodal distribution in $v\sin i$ for their SMC sample as predicted by simulations that include the impacts of binary interactions \citep{Demink2013}. We do not have sufficient statistics to comment conclusively on the particular shape of our distribution, but we do see a small population of stars (approximately a quarter of our sample) with values of $v\sin i$ in excess of $\approx$ 200 km s$^{-1}$. All but one of these stars are among the candidate binary systems identified in Section~\ref{subsec:binary_sel}.

The four binary candidate stars (Stars 2, 5, 6, and 7) all have large values of spectroscopic stellar mass ($M_{\star, \textrm{spec}}>50M_\odot$). If these systems are truly unresolved binaries, their measured photometry would be inflated as compared to each individual star's brightness. Potentially unresolved, blended line profiles in the spectra of these binary candidates could be confused for intrinsically broad lines, increasing their measured values of $\log(g)$, which are among the highest we measure in our sample. Given that the calculation of $M_{\star, \textrm{spec}}$ is directly based on both the photometry and measured $\log(g)$, the otherwise surprisingly high values of $M_{\star, \textrm{spec}}$ are consistent with our identification of these stars as candidate binary systems. If these are indeed binary systems, then the true stellar masses of the individual stars are likely lower than the reported values of $M_{\star, \textrm{spec}}$, which we discuss in more detail in Section~\ref{subsec:masscomp}

Aside from the binary candidates, one of the O stars -- Star 1, the O5-type star -- is significantly hotter, more luminous, and more massive than the rest of the stars in the sample, consistent with expectations for such an early-type star. 

\begin{figure*}[t]
\begin{center}
\includegraphics[width=\linewidth,angle=0]{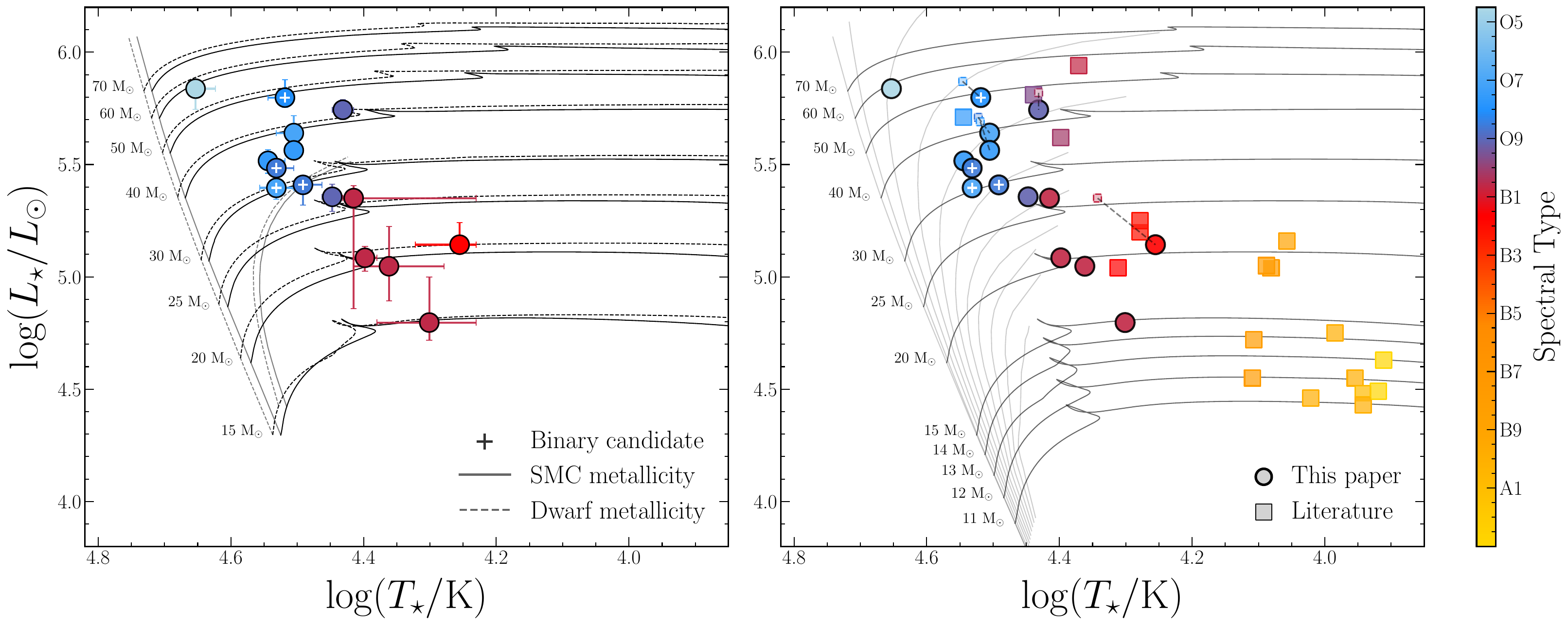}
\caption{(Left) HRD of the OB stars in our sample colored by spectral type with binary candidates marked with a white cross. For some of the stars, the error bars are smaller than the marker. All of the stars are giants or supergiants. The solid black lines show stellar evolutionary tracks from BoOST \citep{Szecsi2022} with SMC metallicity and the dashed black lines show tracks with half-SMC metallicity. The gray vertical lines show isochrones for 0 and 5 Myr with solid lines for SMC metallicity and dashed lines for half SMC metallicity. The lower-metallicity tracks predict slightly older ages and slightly lower masses than the SMC-metallicity tracks. (Right) HRD for the full sample of stars in NGC 3109 with measured parameters. The values from the literature (including from \citet{Evans2007}, \citet{Tramper2014}, and \citet{Hosek2014}) are shown as squares. For stars in our sample with previously published parameters, we indicate the position of the star according to the literature values as a small square, which is connected with a dashed line to the position based on our measured values. The stellar evolution tracks and isochrones in this panel are those with SMC metallicity. The isochrones range from 0--10 Myr and are spaced by 1 Myr. Our sample of OB stars populates the previously sparse young, massive region of the HRD for NGC 3109.} \label{fig:hrd}
\end{center}
\end{figure*}

\subsection{Relation Between $T_\star$ and Spectral Type}\label{subsec:tspec}
While the correlation between effective temperature and spectral type has been well constrained for massive stars in the Milky Way, LMC, and SMC \citep{Massey2004, Massey2005, Trundle2007, Massey2009, SimonDiaz2014, Holgado2018, RamirezAgudelo2017}, it remains uncertain for lower metallicities. Due to their smaller radii, it is expected that stars with lower metallicity will have higher effective temperatures for a given spectral type. \citet{Garcia2013} confirmed this prediction, presenting the first fit to the sub-SMC metallicity spectral type--effective temperature relation, which \citet{Tramper2014} later updated with additional low metallicity stars (including four from NGC 3109). Our sample provides seven additional sub-SMC metallicity O stars with which to further constrain the relation. {The spectral type--temperature relation is known to vary for different LCs \citep{Ramachandran2019}, but due to the lack of observed metal-poor O-dwarf stars (LC IV-V) in our sample and in the larger literature, we focus here on supergiants (LC I), bright-giants (LC II) and giants (LC III).}

Before fitting an updated line to our sample of sub-SMC-metallicity O stars, we first compare our new sources with previously published relations. In the left panel of \autoref{fig:tspec}, we plot the derived values of $T_\star$ versus spectral type along with relations derived for SMC stars of LC I--III \citep{Ramachandran2019} and sub-SMC stars of LC I--III \citep{Tramper2014}. {We also show recently published relations from \citet{Lorenzo2025}, derived from large grids of model spectra of O stars at $0.1\, Z_\odot$ generated with \texttt{FASTWIND} \citep{Puls2005}. Some of the metal-poor O stars in our sample are consistent with the sub-SMC relations and some more closely aligned with the SMC relation. Notably, the O5-type star is significantly hotter than would be predicted for an SMC-metallicity star of its spectral type. The stars in our sample are also largely consistent with the model-derived relations from \citet{Lorenzo2025} for giants and supergiants. The NGC 3109 O stars span LCs I--III, with six LC I, three LC II, and one LC III.}

Next, we fit a line to the relation for the O stars in our sample. {We exclude Star 1 from the fit, because it is an outlier in spectral type within our sample and its inclusion biases the best-fit slope. Without Star 1, we find:}
\begin{equation}
    T_\star \text{ [kK]} = 54.7^{+5.3}_{-5.3}-2.7^{+0.6}_{-0.6} \times \text{ST}.
\end{equation}

where ST is a numerical representation of the spectral type such that an O9 star has ST $ = 9$ and a B1.5 star has ST $=11.5$. {For completeness, we also perform the fit with Star 1, finding a steeper slope and higher temperatures for early spectral types:}
\begin{equation}
    T_\star \text{ [kK]} = 59.9^{+3.5}_{-3.5}-3.3^{+0.4}_{-0.4} \times \text{ST}
\end{equation}

We also compile spectral types and effective temperatures for sub-SMC-metallicity O-giant stars from the samples presented in \citet{Garcia2013} and \citet{Tramper2014}. We do not include the three stars from NGC 3109 presented in \citet{Tramper2014} that overlap with our sample, preferring to use our own temperatures and spectral types for these sources (which are in any case largely consistent). We fit a line to this full sample as well. We derive the relation:
\begin{equation}
    T_\star \text{ [kK]} = 53.7^{+2.5}_{-2.5}-2.4^{+0.2}_{-0.2} \times \text{ST}.
\end{equation}
We show the result of the regressions in the right panel of \autoref{fig:tspec}.

{The regression fit to the combined sample of O stars is slightly lower than the fit from \citet{Tramper2014}. A potential explanation for this discrepancy is that the majority of stars from \citet{Garcia2013} have LC III, but the majority of stars in our NGC 3109 sample have LC I and more evolved stars are expected to have lower temperatures for a given spectral type. This sample difference could also be contributing to the lower normalization of the fit to the NGC 3109 O-giant stars presented in this paper. The fit to the NGC 3109 sample alone (excluding Star 1) is nearly perfectly consistent with the model-derived relation for supergiants from \citet{Lorenzo2025}. Including Star 1 in the fit increases the slope but does not significantly change the overall normalization relative to the literature fits, which would still be lower than previous results from observations for late-type O stars, but consistent with the \citet{Lorenzo2025} supergiant fit. A thorough data-based analysis of the variation in the spectral type--$T_\star$ relation for various LC will require larger samples of metal-poor O star spectra. Given the technical challenges of constructing such a sample, it is promising that the model-based results from \citet{Lorenzo2025} agree well with the observations presented here.} 

\subsection{HR Diagram}\label{subsec:hrd}

Having measured the luminosity and temperature of the stars in our sample, we construct a Hertzsprung-Russell Diagram (HRD) in the left panel of \autoref{fig:hrd} to study the stellar population and evolutionary status of NGC 3109. We also show the stellar evolutionary tracks and isochrones from the Bonn Optimized Stellar Tracks program \citep[BoOST,][]{Szecsi2022} for SMC metallicity (0.2 $Z_\odot$) and half-SMC metallicity (0.1 $Z_\odot$). The gas-phase metallicity of NGC 3109 is in between these two values at $\sim0.12$ $Z_\odot$. The BoOST tracks are initialized with rotation speeds of 100 km s$^{-1}$, similar to the average $v\sin i$ of our sample. 

We determine the initial mass $M_0$ and evolutionary mass $M_\text{evo}$ of each star by identifying the evolutionary track which is closest to the star's location in the HRD. $M_\text{evo}$ is taken to be the mass of the evolutionary track at the position closest to the star's current position and $M_0$ is taken to be the initial mass of the track. We note that while we show only a small selection of tracks in \autoref{fig:hrd}, there are over 1,800 tracks for each metallicity and so the masses are determined with precision ($\sigma_{M_\text{evo}}\approx 5\, M_\odot$). The values of $M_{\text{evo}}$ and $M_0$ are reported for the SMC tracks in \autoref{tab:stellarparams}. The evolutionary tracks with 0.1\,$Z_\odot$ yield initial and evolutionary masses that are on average 1\,$M_\odot$ lower than the tracks with SMC metallicity. We report the values derived from the SMC-metallicity tracks to be consistent with the SMC-metallicity PoWR atmosphere models that we use in Section~\ref{sec:fitting} to fit for the stellar parameters $L_\star$ and $T_\star$.

The O and B stars in our sample fall in the expected region of the HRD, with the earliest O stars being the hottest and most luminous. The evolutionary masses of the O stars range from $25\,M_\odot -60\,M_\odot$ and the B-star masses range from $15\,M_\odot -25\,M_\odot$. The O stars are all located between the zero-age main sequence (ZAMS) and terminal-age main sequence (TAMS), while the B stars appear to be somewhat more evolved, consistent with their supergiant LC. The four stars we have identified as candidate binary systems are marked with white crosses in \autoref{fig:hrd} and we note that one of these stars (Star 6) is among the most luminous in the sample. Its position in the HRD (and the positions of the other binary candidates) could be impacted by a potentially unresolved companion. 

\begin{figure}[t]
\begin{center}
\includegraphics[width=\linewidth,angle=0]{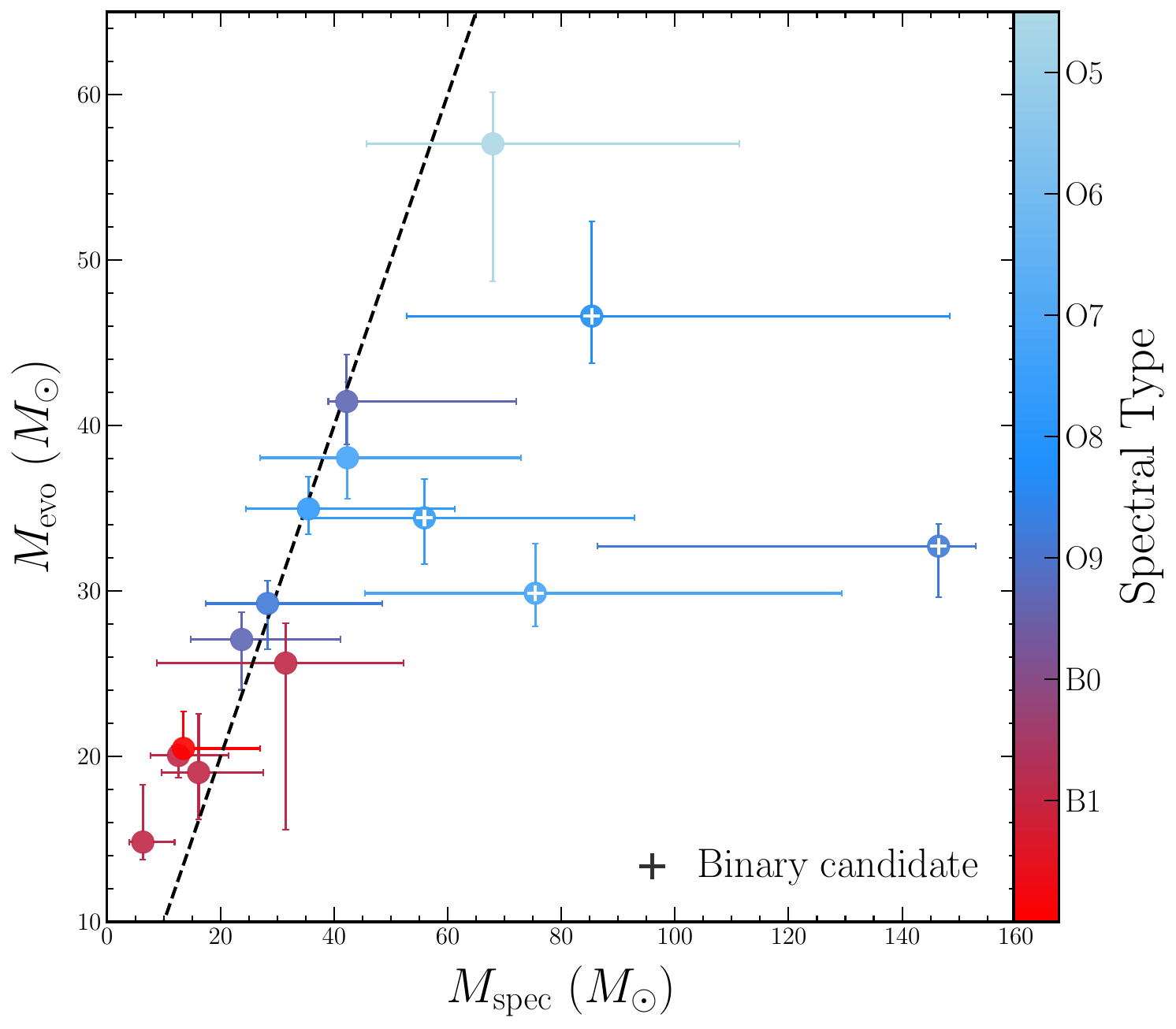}
\caption{Evolutionary mass versus spectroscopic mass for the OB stars in our sample, colored by spectral type. The one-to-one line is shown as a black dashed line and the binary candidates are marked with white crosses. The four O stars with $M_\text{spec}>>M_\text{evo}$ are the stars we have identified as binary candidates. The remaining O-star masses are in reasonable agreement, but the B-star spectroscopic masses are systematically slightly smaller than the evolutionary masses for all but one B star.}\label{fig:masscomp}
\end{center}
\end{figure}
\begin{table*}
    \caption{Measured stellar parameters compared to stellar parameters from the literature. \label{tab:litcomp}}
    
\centering
\begin{tabular}{ccccccccccccc}
\hline\hline
     \multicolumn{7}{c}{This Paper} &  \multicolumn{6}{c}{Literature}\\
     \cmidrule(lr){1-7}\cmidrule(lr){8-13}
     Star ID &
     T & $\log g$ & $v\sin i$ & $R_\star$ & $\log(L_\star)$ & M$_\star$ & 
    T & $\log g$ & $v\sin i$ & $R_\star$ & $\log(L_\star)$ & M$_\star$ \\   
     &(kK) & (cm s$^{-2}$) & (km s$^{-1}$) & ($R_\odot$) & ($L_\odot$) & (M$_\odot$) &(kK) & (cm s$^{-2}$) & (km s$^{-1}$) & ($R_\odot$) & ($L_\odot$) & (M$_\odot$)\\
(1) & (2) & (3) & (4) & (5) & (6) & (7) & (8) & (9) & (10) & (11) & (12) & (13)\\
     \cmidrule(lr){1-7}\cmidrule(lr){8-13}
3 & 	32$_{-1}^{+2}$ & 	3.4$_{-0.2}^{+0.2}$ & 	160$_{-40}^{+40}$ & 	21$_{-1}^{+1}$ & 	5.64$_{-0.06}^{+0.08}$ & 	42$_{-16}^{+31}$  & 33.30 & 3.35 & 200 & 21.9 & 5.71 & 38.9 \\
4 & 	32$_{-1}^{+1}$ & 	3.4$_{-0.2}^{+0.2}$ & 	80$_{-40}^{+40}$ & 	20$_{-1}^{+1}$ & 	5.56$_{-0.03}^{+0.05}$ & 	35$_{-11}^{+26}$      & 33.05 & 3.16 & 82  & 21.8 & 5.69 & 25.0 \\
6 & 	33$_{-1}^{+2}$ & 	3.6$_{-0.2}^{+0.2}$ & 	120$_{-40}^{+40}$ & 	24$_{-1}^{+2}$ & 	5.80$_{-0.05}^{+0.08}$ & 	85$_{-33}^{+64}$ & 35.15 & 3.53 & 110 & 23.7 & 5.87 & 69.1 \\
9 & 	27$_{-1}^{+1}$ & 	3.0$_{-0.0}^{+0.2}$ & 	80$_{-40}^{+40}$ & 	34$_{-2}^{+2}$ & 	5.74$_{-0.05}^{+0.05}$ & 	42$_{-4}^{+30}$    & 27.0  & 2.90 &  –– & 37   & 5.82 & 39.7 \\
17 & 	18$_{-1}^{+3}$ & 	2.4$_{-0.0}^{+0.2}$ & 	80$_{-40}^{+40}$ & 	38$_{-7}^{+5}$ & 	5.14$_{-0.02}^{+0.10}$ & 	13$_{-2}^{+14}$   & 22.0 & 2.60 &  ––  & 33 & 5.35 & 15.6 \\
\end{tabular}

    \tablecomments{Literature values for Stars 3, 4, and 6 are taken from \citet{Tramper2014} and values for Stars 9 and 17 are taken from \citet{Evans2007}. Column (1): the identification number assigned to each star. Column (2): stellar temperature. Column (3): log surface gravity. Column (4): projected rotation speed. Column (5): stellar radius. Column (6): stellar luminosity. Column (7): spectroscopic stellar mass. Column (8): literature stellar temperature. Column (9): literature log surface gravity. Column (10): literature projected rotation speed. Column (11): literature stellar radius. Column (12): literature stellar luminosity. Column (13): literature spectroscopic stellar mass.}
\end{table*}

In the right panel of \autoref{fig:hrd} we compile stars in NGC 3109 from the literature with spectroscopically measured parameters. The full sample includes a total of 34 stars: the ten O stars and five B stars in our sample with well-measured parameters, six B-type supergiants from \citet{Evans2007} (excluding the two which overlap with our sample), one O-type {supergiant} from \citet{Tramper2014} (excluding the three which overlap with our sample), and five late B-type {supergiants} and seven early A-type {supergiants} from \citet{Hosek2014}, which are cooler and more evolved than the earlier-type stars in the sample. We also show the positions of stars whose parameters had been previously measured, but have been reclassified and measured using our new spectra. 

The spread in temperature and evolutionary ages of this large sample of field stars suggests that there has been extended star formation in NGC 3109. The presence of cooler, evolved A stars is evidence for star formation over $\sim25$ Myr ago. Star 1, which we have characterized here for the first time, is the hottest, youngest, and most massive star so far measured in NGC 3109. Its position in the HRD provides evidence of recent star formation in the galaxy -- at least as recently as 2 Myr ago. 

\subsection{Comparing Spectroscopic and Evolutionary Masses}\label{subsec:masscomp}
In Section~\ref{subsec:hrd}, we estimated $M_\text{evo}$ and $M_0$ from the BoOST stellar evolution tracks for SMC metallicity. In \autoref{fig:masscomp}, we compare these evolutionary masses with the spectroscopic masses we derived in Section~\ref{subsec:fitting_phot}.

While the errors on $M_{\star, \text{spec}}$ are large and the spectroscopic and evolutionary masses of most stars are consistent within the uncertainty, there are still a number of interesting results from this comparison. Firstly, the four stars which have $M_{\star, \text{spec}} >> M_\text{evo}$ are those stars which we identified as binary candidates based on the discrepancies between their observed photometry and the SEDs predicted by their best-fit spectroscopic models. If they are indeed binary systems, then the observed luminosity, mass, and radius would be inflated from what would be measured from a single star and would not match predictions based on spectroscopically-derived values of temperature and surface gravity. Others have argued previously that systematically higher spectroscopic masses as compared to evolutionary masses are a result of binarity \citep{Bouret2013}.

The only other notable mass discrepancy for the O stars is a slightly higher $M_{\star, \text{spec}}$ for the earliest O star, Star 1. We estimate $M_\textrm{evo}=57\,M_\odot$, lower than the measured $M_{\star, \textrm{spec}}=68\, M_\odot$ for this star. Regardless, both the spectroscopic and evolutionary masses we derive for this star make it the most massive apparently single O star identified in NGC~3109, and place it among the highest-mass O stars known at sub-SMC metallicity to date. 

The evolutionary and spectroscopic mass measurements of the remaining five O stars agree well. \citet{Ramachandran2018b} found that for stars in the LMC, while spectroscopic masses for dwarf O stars are systematically higher than their evolutionary masses, the bias is much smaller for more evolved stars. Given that all of our O stars are giants or supergiants, the general mass agreement is in line with expectations for these stars.

Although $M_{\star, \text{spec}}$ and $M_{\text{evo}}$ are statistically consistent for all but one of the lower-mass stars, $M_{\star, \text{spec}}$ is systematically lower than $M_{\text{evo}}$ for the stars with $M_{\star, \textrm{spec}}<20 M_\odot$. \citet{Nieva2014} found that the bias towards higher $M_{\star, \text{spec}}$ applies to stars with $M_{\star, \textrm{spec}} \gtrsim 15 M_\odot$ and below that, the two masses are in general agreement. Our small sample size prevents us from commenting strongly on this question, but we note that the specific values of $M_\textrm{evo}$ depend on the evolutionary tracks and their chosen stellar physics models, in particular for mass-loss rates and rotation speed. Given that the systematic difference we observe is quite small, it is possible that a different choice of evolutionary tracks would erase this bias. In fact, as we mentioned in Section~\ref{subsec:hrd}, the tracks with $Z=0.1\,Z_\odot$ have $M_\text{evo}$ that are $\sim1\,M_\odot$ less than the SMC metallicity tracks; using these values would also lessen the observed bias.

\section{Discussion}\label{sec:disc}

\begin{figure*}[t]
\begin{center}
\includegraphics[width=\linewidth,angle=0]{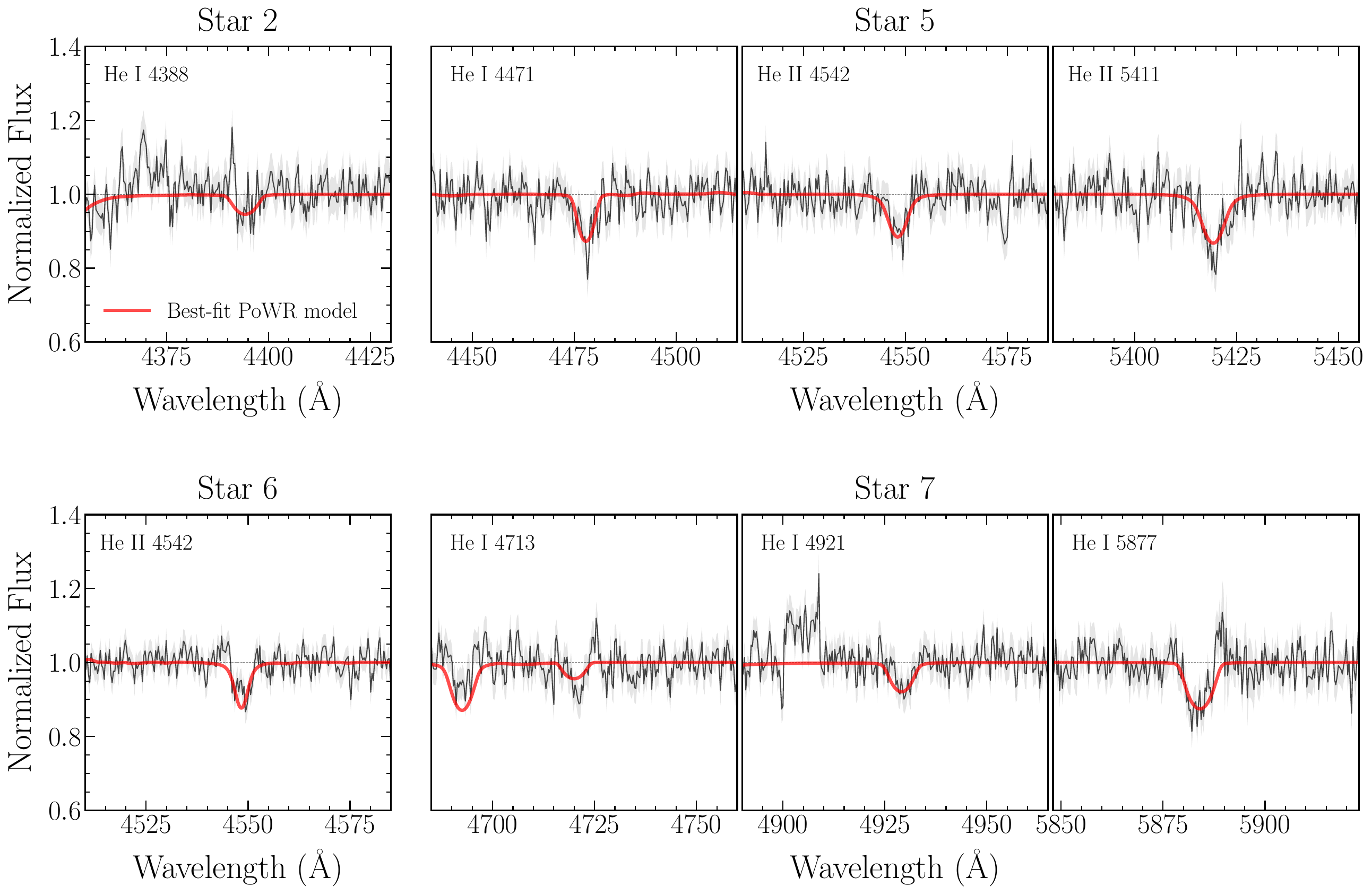}
\caption{Examples of line profiles in the candidate binary stars that are not well fit by the optimal OB-star PoWR models. The spectra are shown in black with the one $\sigma$ uncertainty shaded in gray. The best-fit PoWR model for each star is overlaid on the spectrum in red. While the resolution and SNR of the spectra make it challenging to conclusively identify blended lines from two superimposed stellar spectra, it is notable that the stars selected as binary candidates based on the disagreement between their observed and predicted SEDs also exhibit some irregular line profiles. \label{fig:binarylines}}
\end{center}
\end{figure*}
\subsection{A Large Sample of Massive Stars in NGC 3109}\label{sec:stellarparams}

Though ours is certainly not the first spectroscopic study of massive stars in NGC 3109, our sample represents a significant addition to the massive star literature in this galaxy. \citet{Evans2007} presented low-resolution optical spectroscopy ($R\sim1000$) of 91 massive O-, B-, and A-type stars stars in NGC 3109, but determined stellar parameters for only eight of the early B-type supergiants using stellar atmosphere models from \texttt{FASTWIND} \citep{Puls2005}. \citet{Tramper2014} analysed medium-resolution optical spectra ($R\sim$ 7000, SNR\,$\sim$\,25--45) of four O stars from the 91 stars from \citet{Evans2007} also using models from \texttt{FASTWIND}. \citet{Hosek2014} reanalyzed a number of the spectra from \citet{Evans2007}, but focused on A-type stars and did not contribute any new stellar parameters for O or early B-type stars. In total, stellar parameters have been previously published for only four O stars and eight early-B stars in NGC 3109.

Our sample contributes stellar parameters for ten O stars and five B stars, including six O stars and four B stars with no previously published parameters. Our results for these ten new OB stars nearly double the number of massive stars in NGC 3109 with measured stellar parameters. Five of the stars in our sample (four O stars and one B star) overlap with those analyzed in \citet{Evans2007} and \citet{Tramper2014}. We compare our measured values with the parameters published in these studies in \autoref{tab:litcomp}; they are largely consistent within the uncertainties, but we do measure slightly higher surface gravities, lower luminosities, and larger masses than \citet{Tramper2014} for our overlapping stars. Of the remaining 12 stars in our sample (including stars 11 and 15 for which we do not report measured parameters), seven were classified but not modeled in \citet{Evans2007} using the low resolution optical spectroscopy and as far as we are aware, there is no existing spectroscopy or published analysis of the remaining five stars in our sample. 

Including our results, there are now 22 OB stars in NGC 3109 with published stellar parameters. This is the largest existing sample of OB stars for any sub-SMC metallicity galaxy. The NGC 3109 sample is a uniquely powerful dataset that will enable future constraints on models of low-metallicity stellar evolution.

\subsection{Binarity at Low Metallicity}\label{subsubsec:binary}

It is well established that interactions with stellar companions are both common and highly influential in the evolution of massive stars, at least at Galactic metallicity \citep{Kobulnicky2007, Mason2009, Sana2012}. It is less understood how frequently stellar binarity or multiplicity occurs for metal-poor massive stars, although there is strong evidence that the binary fraction of solar-mass stars increases with decreasing metallicity \citep{Moe2019}. Spectroscopic surveys in the LMC have measured OB binary fractions comparable to those found in the Milky Way \citep[50\% -- 70\%;][]{Kiminki2012, Sana2013,Dunstall2015, Guo2022, Banyard2022} and a large campaign is underway to characterize massive-star binarity in the SMC \citep[BLOeM;][]{Shenar2024}. Despite these considerable efforts, multiplicity of massive stars remains largely unexplored in sub-SMC metallicity environments. A comprehensive study in these environments would require repeated spectroscopic measurements of a large sample of stars in local low-$Z$ dwarf galaxies -- an expensive undertaking. 

Although the moderate resolution and single-epoch spectroscopy in this work does not allow for the conclusive characterization of binarity in our sample, we do find potential evidence of binarity in four of the O stars and present them as binary candidates. In Section~\ref{subsec:binary_sel}, when comparing the model SEDs predicted by the spectroscopically-fitted parameters of Stars 2, 5, 6, and 7 to their observed photometry, we found that each of these stars was significantly brighter than predicted by factors of 4, 2, 3, and 9 respectively. The extreme factor of 9 discrepancy in Star 7 could be attributed to the unresolved superposition of Balmer lines, which would artificially broaden the lines and inflate the measured value of $\log(g)$, reducing the predicted luminosity. Higher-resolution or multi-epoch spectroscopy would be needed to verify this hypothesis.

These binary-candidate stars also have the largest discrepancy between their $M_{\star, \textrm{spec}}$ and $M_\textrm{evo}$ (see \autoref{fig:masscomp}). This discrepancy could be explained by the presence of an unresolved companion, which would boost the measured photometry and therefore the spectroscopic mass. We do note that $M_{\star,\textrm{spec}}$ and the multiplicative model-to-data photometric factor $c$ have high uncertainties (see \autoref{tab:stellarparams}), but we find additional evidence supporting the binary candidate designation of these stars.  

Stars 2, 5, and 7 are among the fastest rotators in the sample with measured $v\sin i$ values of 280, 200, and 240 km s$^{-1}$, respectively. Projected stellar rotation speeds greater than $\approx 200$ km s$^{-1}$ are an expected result of binary interactions \citep{Demink2013} and the superposition of absorption lines can cause the appearance of broadening, leading to a higher preferred value of $v\sin i$. We also observe that some of the absorption lines in the spectra of these stars appear irregular and are not fit well by the models as seen in \autoref{fig:binarylines}. 

Given the moderate resolution and low SN of the data, we cannot conclusively say whether these profiles are the result of binarity. Nevertheless, we believe that the combined evidence presented above demonstrates that these four stars are optimal candidates for future spectroscopic monitoring. If the binary classifications are substantiated, our sample would have a binary fraction of 24\% overall or 40\% for the O stars.

\subsection{Low-metallicity Stars with High Mass Loss}\label{subsec:massloss}

The nature of stellar mass loss at low metallicity is highly consequential for evolutionary models of massive stars and their ionizing flux, yet these low-$Z$ models remain almost entirely unconstrained by observations \citep{Vink2001, Krticka2018, Bjorklund2021, Vink2021, GormazMatamala2024}. The strength of radiation-driven winds is tied directly to metal-line opacity, and is therefore expected to be weaker in metal-poor stars. Even so, mass-loss rates predicted by models of SMC-metallicity massive stars disagree dramatically with observations \citep{Martins2004, Martins2005, Marcolino2009, Bouret2013, Ramachandran2019}.

While a quantitative study of mass-loss rates in these metal-poor OB stars would require FUV spectroscopy, we identify \trans{he}{ii}{4686} emission in the optical spectra of two of the O stars in our sample: Stars 1 and 3. Emission in the \trans{he}{ii}{4686} line is only seen in evolved giant and supergiant stars and is used as an indicator of strong stellar winds and mass loss. The emission is of moderate strength in Star 3 ($\textrm{EW} = 0.26$\,\AA) and of exceptional strength for a metal-poor O star in Star 1 ($\textrm{EW} = 1.74$\,\AA).

\begin{figure*}[ht]
\begin{center}
\includegraphics[width=\linewidth,angle=0]{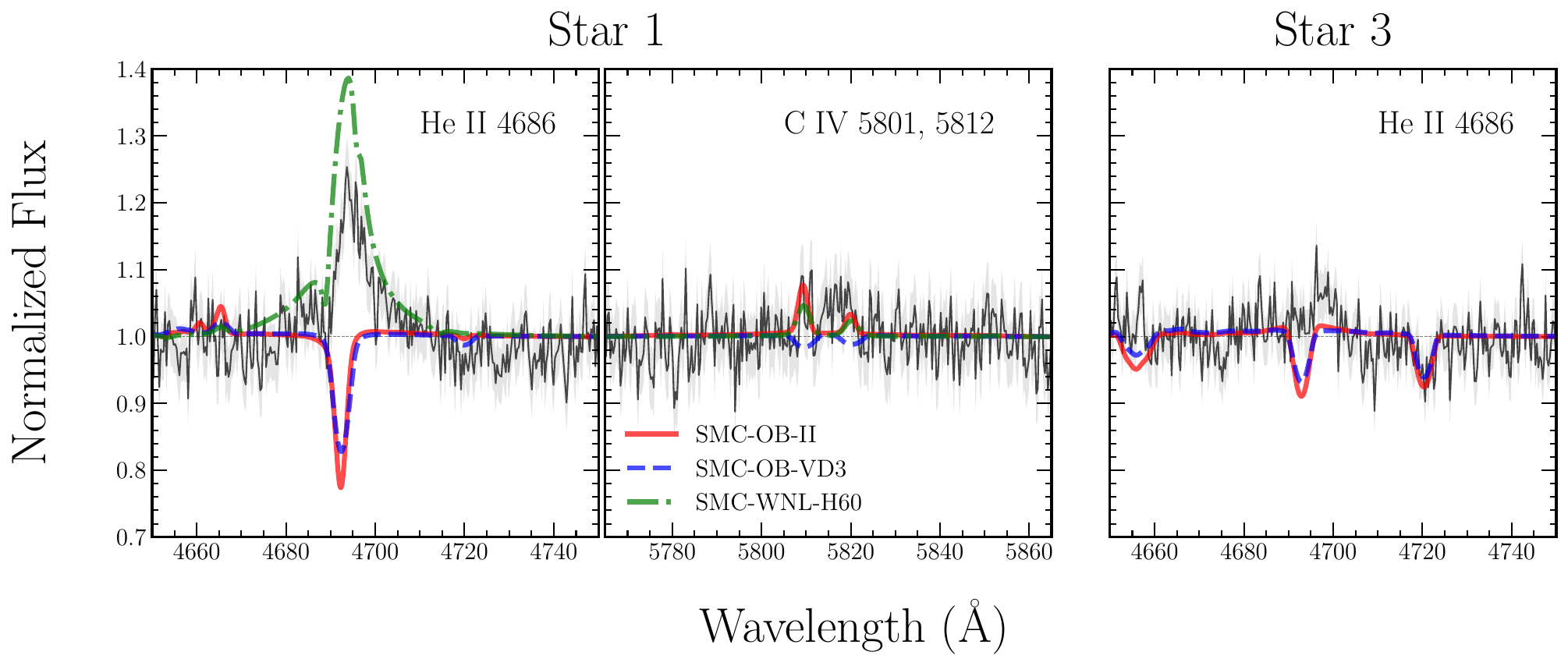}
\caption{Evidence of strong mass loss in Stars 1 and 3. The left two panels show \trans{he}{ii}{4686} emission and \ion{C}{4} $\lambda$5801 and $\lambda$5812 emission in the optical spectrum of Star 1. The gray shaded region indicates the 1$\sigma$ error on the continuum-normalized spectrum. The blue and red lines show the best-fit model spectrum from the default (SMC-OB-VD3) and high-mass-loss (SMC-OB-II) PoWR grids. The green dash-dotted line shows the PoWR WR model that best matches the observed \trans{he}{ii}{4686} emission in Star 1 as described in Section~\ref{subsec:massloss}. The right-most panel shows the \trans{he}{ii}{4686} emission in Star 3 with its best PoWR model fits from the default and high-mass-loss grids. None of the OB PoWR models fit the \trans{he}{ii}{4686} emission well, but the high-mass-loss model better matches the \ion{C}{4} emission in Star 1. The WR model matches the observed profile of the \trans{he}{ii}{4686} emission in Star 1, but overpredicts the strength of the emission. \label{fig:mass_loss}}
\end{center}
\end{figure*}

 As mentioned in Section~\ref{subsec:fitting_spec}, we employ the high-mass-loss grid, SMC-OB-II, for fitting the spectra and photometry of both of these stars because they exhibit \trans{he}{ii}{4686} emission (see \autoref{fig:mass_loss}). Comparing the best-fit models to the optical spectra in \autoref{fig:mass_loss} shows that neither of the PoWR O-star models fit the \trans{he}{ii}{4686} emission remotely well. In fact none of the model spectra in either PoWR grid shows \trans{he}{ii}{4686} emission. (Although in Star 1, we also see evidence of \ion{C}{4} $\lambda$5801 and $\lambda$5812 emission, which is fit better by the high-mass-loss grid than by the default grid.) Although it is possible to create a PoWR model with \trans{he}{ii}{4686} emission by fine-tuning the wind parameters, the issue of accurate \trans{he}{ii}{4686} modeling is a well-known problem.

\citet{Sander2024} performed a comprehensive comparison of various spectral analysis methods (including some using PoWR models) and found that nearly all of the models fail to accurately fit the \trans{he}{ii}{4686} line, although a number of the models they consider do produce emission. \citet{Martins2021} generated synthetic spectra for the MS to post-MS evolutionary stages for massive stars with SMC-metallicity and with $Z_\odot/30$ metallicity, including for a $60\,M_\odot$ star. None of their spectra for this star exhibit such strong \trans{he}{ii}{4686} emission as we observe in Star 1, which is telling given that their metallicity values bracket that of NGC 3109 and the stellar mass of $60\,M_\odot$ is very close to that of Star 1, which has $M_{\star, \textrm{spec}}=68\,M_\odot$ and $M_\text{evo} = 56\,M_\odot$.

\begin{figure*}[ht]
\begin{center}
\includegraphics[width=\linewidth,angle=0]{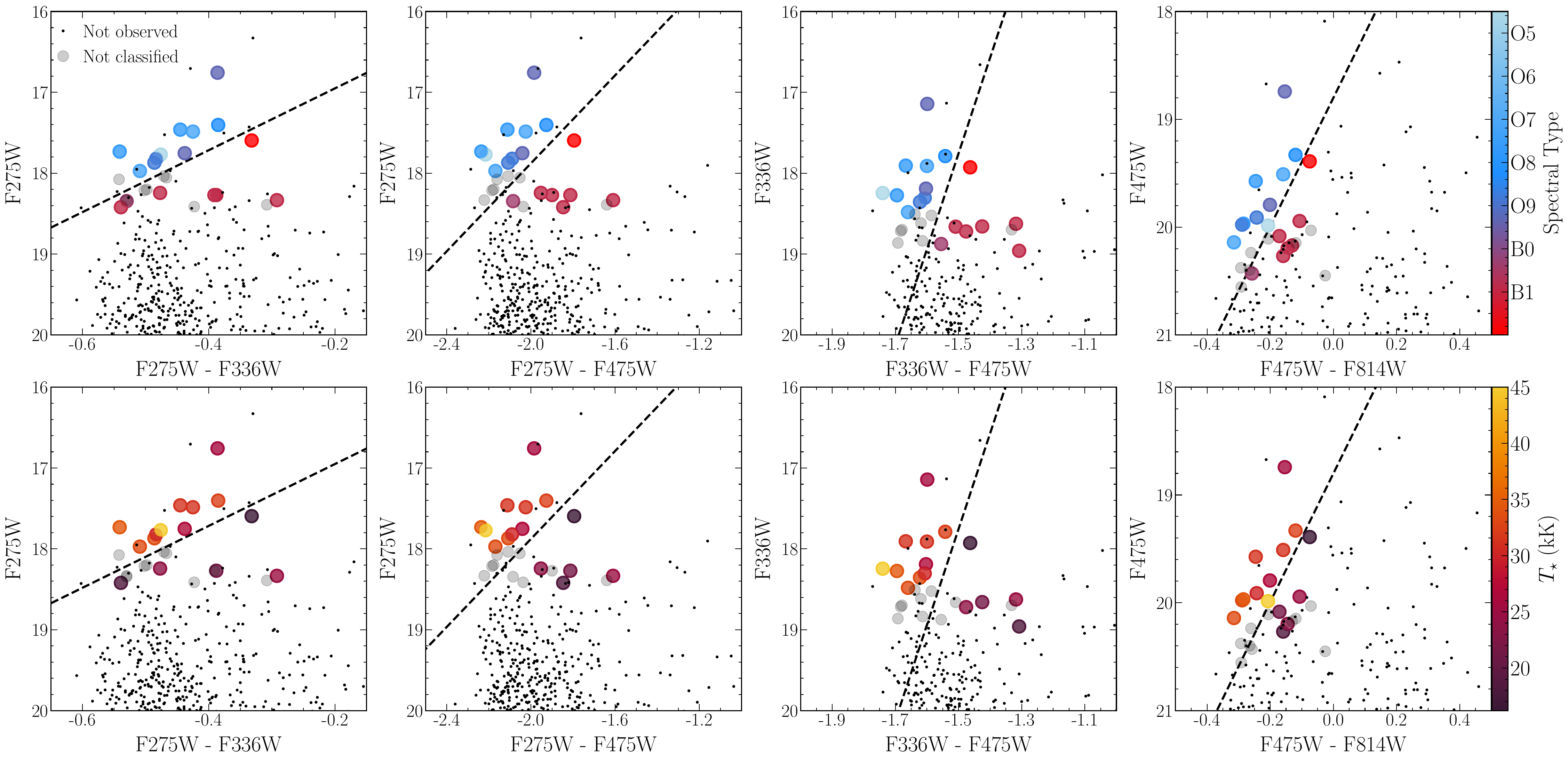}
\caption{Color-magnitude diagrams of stars in NGC 3109 made with HST optical and NUV photometry. In the top row, the stars are colored by spectral type and in the bottom row by $T_\star$. We show the linear boundaries we derive in Section~\ref{subsec:select} to separate the O stars from the B stars as dashed black lines. While all of the derived boundaries successfully separate the classes of stars, those that include NUV photometry are more useful for identifying the hottest stars. \label{fig:big_cmd}}
\end{center}
\end{figure*}

We also consider the possibility that Star 1 is a rare, low-metallicity Wolf-Rayet (WR) star, although confirming the star's evolutionary status is challenging given our limited wavelength coverage and the high uncertainty in \trans{he}{ii}{4686} modeling. WR stars are massive stars ($M_\star>25\,M_\odot$) in the final stages of evolution that are characterized spectroscopically by strong, broad emission lines -- in particular \trans{he}{ii}{4686} and often carbon and nitrogen emission \citep{Crowther2007}. The populations of Galactic WR stars and WR stars in the Magallenic Clouds have been thoroughly characterized \citep[e.g.,][]{Massey2014, Massey2015,Rosslowe2015, Massey2017, Neugent2018, Crowther2023}, but observations of WR stars at sub-SMC metallicity are exceedingly rare. To be spectroscopically classified as a WR (at least at SMC metallicity and above), a star should have \trans{he}{ii}{4686} emission with EW of at least 10\,\AA\ with the peak of the emission reaching at least twice the level of the continuum and should exhibit emission in the H$\beta$ line \citep{Massey2014, Shenar2020}. Star 1 meets none of these criteria, and so we are confident in its classification as an O5 If star. 

Because the spectral appearance of metal-poor WR stars is highly uncertain, we also compare the spectrum of Star 1 to model WR spectra from PoWR. In addition to the OB-star grids, precomputed grids of WR stars with varying metallicity and hydrogen fraction are publicly available. Nearly all of these models predict strong, broad line emission that is clearly inconsistent with the spectrum of Star 1. A small number of models, including an SMC metallicity model with a 60\% hydrogen mass fraction (\texttt{smc-wnl-h60\_08-02} as shown in \autoref{fig:mass_loss}) have spectra that are reasonably consistent with the optical spectrum of Star 1. Even this model overpredicts the strength of \trans{he}{ii}{4686} and it occupies an extreme region of the WR model parameter space that is poorly covered by the grid; the value of $R_t=1.9$ (transformed radius, a function of mass-loss rate and stellar radius used to parameterize the WR models) for this model is high. \citet{Shenar2020} demonstrated that stars with $R_t$ above a certain temperature-dependent threshold would not appear as WR stars and would be classified as Of stars, just as Star 1 is classified. The only significant difference in the optical spectrum of the best WR model and the best-fit O-star model is the \trans{he}{ii}{4686} emission, which as we described above is challenging to model accurately -- especially at low metallicity. Based only on the optical spectroscopy, it is therefore not possible to determine with confidence which of these models is truly a better fit and so we cannot determine the evolutionary status of Star 1 with certainty.

Whether Star 1 is a young O star with unexpectedly high mass loss, or an evolved WR star with unexpectedly low mass loss, the \trans{he}{ii}{4686} emission in Star 1 is indeed quite surprising for such low metallicity. Additional constraints on the wind-strength and mass-loss rates of this star are necessary, and will be enabled with planned future observations of its FUV spectrum (HST-GO-17491; PI: O.\ G.\ Telford). Confirmation of low stellar metallicities and strong mass loss in Stars 1 and 3 would be somewhat unexpected and could provide a new constraint for low-$Z$ massive stellar wind theory. It is also possible that the abundances of these stars are higher than the gas-phase abundance of NGC 3109 would suggest, which would bring them closer in line with current theoretical predictions.

Lastly, we note that \citet{Tramper2011} and \citet{Tramper2014} derived a high mass loss rate ($\dot{M} = 10^{-5.35}\,M_\odot$\,yr$^{-1}$) for Star 6 -- based on the completely filled in \trans{he}{ii}{4686} line and the weak H$\alpha$ absorption in their optical spectra.  Our optical spectrum of the same star does not cover H$\alpha$, but we find \trans{he}{ii}{4686} in absorption and no additional evidence of strong mass loss. We do identify this star as a binary candidate in Section~\ref{subsubsec:binary}, so it is possible that this disagreement is a result of changing line profiles due to the binarity of the system. It could also be demonstrative of the uncertainty associated with mass-loss measurements derived purely from optical data \citep[e.g.,][]{Bouret2015}.

\subsection{Lessons for Selecting Metal-poor O Stars from NUV Photometry}\label{subsec:select}

One of the main challenges in observing, classifying, and characterizing O stars is identifying likely candidates from photometry. While candidates can be selected using easily accessible ground-based optical photometry, NUV photometry is ideal for identifying O stars, whose SEDs peak in the UV. Our selection criteria based on HST NUV photometry described in Section~\ref{sec:data} and shown in \autoref{fig:selection_cmd} proved to be quite successful in identifying young, early-type massive stars. Of the 17 stars that we classified, ten were classified as O stars and the remaining seven as Early-B stars. 

To better inform future metal-poor O-star candidate selection, we construct CMDs of the classified stars in our sample and fit linear boundaries separating the O and B stars using a support vector machine (SVM) algorithm. SVM algorithms are used to obtain classification boundaries by maximizing the distance between the boundary and closest members of each class in the given parameter space. The O and B stars in our sample are easily linearly separated in the CMDs, so this algorithm is well suited to our data. We derive the following selection criteria for O stars:
\begin{equation}
    \textrm{F275W} < -3.83 \times (\textrm{F275W}-\textrm{F336W}) +16.19
\end{equation}
\begin{equation}
    \textrm{F275W} < -2.72 \times (\textrm{F275W}-\textrm{F475W}) +12.43
\end{equation}
\begin{equation}
    \textrm{F336W} < -11.84 \times (\textrm{F336W}-\textrm{F475W}) + 0.00
\end{equation}
\begin{equation}
    \textrm{F475W} < -5.96 \times (\textrm{F475W}-\textrm{F814W}) +18.80
\end{equation}
We show the CMDs for various color-magnitude combinations in \autoref{fig:big_cmd} colored by both spectral type and $T_\star$ along with the derived linear boundaries. While the linear boundaries are able to separate O and B stars in all of the color-magnitude combinations we present, those that include NUV photometry best select for the hottest stars. In the optical CMD, F475W vs. F475W--F814W, the earliest O star (Star 1) is barely above the O-B boundary, but it is easily selected in F336W vs. F336W--F475W. We therefore recommend that future studies interested in selecting for the earliest-type, hottest stars use NUV photometry where available. Based on our derived boundaries and the HST NUV photometry of NGC 3109, there are nearly a dozen additional stars in the galaxy that are O-star candidates, including many of the stars for which we have spectra that are too noisy to be useful for quantitative analysis. Follow-up observations of these candidates with similar medium resolution optical spectroscopy could help expand the size of the NGC 3109 massive-star sample.

We note that the NGC 3109 stars are minimally affected by reddening internal to the galaxy; the median $E(B-V)_\textrm{int}$ is 0.005. The selection criteria we present could be less successful for targets with high extinction.

\section{Summary}
We have presented new medium-resolution optical spectroscopy for 25 OB stars in the dwarf galaxy NGC 3109. We classify 17 of the stars in the sample whose spectra have sufficiently high signal-to-noise. We analyze the spectra and photometry of 15 of these stars, fitting model spectra and SEDs from the publicly available PoWR grids to determine $T_\star$, $\log$(g), $v\sin i$, $R_\star$, $M_\star$, and $L_\star$. Our catalog contributes 15 OB stars with measured stellar parameters, ten of which have not had their parameters previously published and four of which had no preexisting optical spectroscopy of any resolution. 

Using our new catalog, we examine how our OB stars align with previous results for the sub-SMC spectral type-temperature relation. {We find that our O stars are cooler than would be predicted by previous sub-SMC relations, potentially because they are more evolved than O stars in previously used samples. The fit to our sample excluding Star 1 is in perfect agreement with the model-derived relation for supergiants with $0.1\,Z_\odot$ from \citet{Lorenzo2025}.}

We construct an HRD of the stars and find that our stars fill in a previously sparse region of the NGC 3109 HRD -- less-evolved, luminous blue O stars. The earliest-type star in our sample is the hottest, youngest, and most massive star so far confirmed in NGC 3109 and among the most massive stars at sub-SMC metallicity. It lies on the 2 Myr isochrone on the HRD, providing evidence that star formation was ongoing in NGC 3109 even more recently than previously thought.

Additionally, we identify four candidate binary stars based on the strong discrepancy between their predicted SEDs and their observed photometry. While the resolution and SN of our spectra are not high enough to conclusively classify the binary status of these stars, there are a number of oddly shaped line profiles that are not well fit by the models and three of these stars have high values of $v\sin i$. 

Two other stars in our sample exhibit unexpected evidence of strong mass loss; they show emission in \trans{he}{ii}{4686} -- in the case of Star 1, exceptionally strong emission -- that is not well fit by any of the models. We confirm that this star is not a WR star according to conventional classification criteria, but its high mass loss and young evolutionary status make it extremely unique. 

This star and the entire sample provide new and crucial constraints for models of metal-poor stellar evolution and will be combined with future FUV spectroscopy to further specify this information with measurements of mass-loss rates. Theories of massive star evolution, mass loss, and ionizing photon production at low metallicity remain extremely poorly constrained empirically, but our new catalog of stellar parameters for a significant number of metal-poor OB stars substantially increases the available observational data. 
\\
\\

A.M. acknowledges support from the National Science Foundation Graduate Research Fellowship under Grant No. 2039656.
O.G.T. acknowledges support from a Carnegie-Princeton Fellowship through Princeton University and the Carnegie Observatories.

This work was supported by a NASA Keck PI Data Award, administered by the NASA Exoplanet Science Institute. Data presented herein were obtained at the W. M. Keck Observatory from telescope time allocated to the National Aeronautics and Space Administration through the agency's scientific partnership with the California Institute of Technology and the University of California. The Observatory was made possible by the generous financial support of the W. M. Keck Foundation.
The authors wish to recognize and acknowledge the very significant cultural role and reverence that the summit of Maunakea has always had within the indigenous Hawaiian community. We are fortunate to have the opportunity to conduct observations from this mountain.

Based on observations with the NASA/ESA Hubble Space Telescope obtained at the Space Telescope Science Institute, which is operated by the Association of Universities for Research in Astronomy, Incorporated, under NASA contract NAS5-26555. 

This research used NASA's Astrophysics Data System and the arXiv preprint server. 

\facilities{Keck:II (DEIMOS), HST (WFC3)}

\textit{Software}: \texttt{NumPy} \citep[\url{https://numpy.org},][]{Harris2020}, \texttt{Astropy}  \citep[\url{https://astropy.org},][]{Astropy2013, Astropy2018, Astropy2022}, \texttt{Matplotlib} \citep[\url{https://matplotlib.org},][]{Hunter2007}, \texttt{SciPy} \citep[\url{https://scipy.org},][]{Virtanen2020}, \texttt{dsimulator} \citep[\url{https://www2.keck.hawaii.edu/inst/deimos/dsim.html}][]{}, \texttt{spec2d} \citep{Cooper2012, Newman2013}, \texttt{dustmaps} \citep{Green2018}, \texttt{DOLPHOT} \citep{Dolphin2000, Dolphin2016}.

\appendix

\section{NUV and Optical Photometry from HST}\label{app:phot}

We present the NUV and optical HST photometry of the sources we observed with Keck/DEIMOS in \autoref{tab:phot}. Details on the photometry are provided in Section~\ref{sec:data}. 

\section{Additional comments on spectral classification for each star}\label{app:class}

Here we provide additional details on the classification of the OB stars in our sample. As described in Section~\ref{sec:spectralclass}, we rely on classification criteria presented in \citet{Mathys1988, Sota2011, Sota2014, Martins2018, Evans2015}. The assignment of subtypes for the O stars is based on the relative strengths of \hei\ to \heii\ absorption lines in the blue optical wavelength range of the spectra. We report the equivalent widths of the relevant lines in \autoref{tab:ews}.

\underline{\textbf{Star 1  (O5 If):}} Star 1 shows evidence of strong and significant \heii\ absorption and so is classified as an O star. Based on the ratio of \trans{he}{i}{4471}/\trans{he}{ii}{4542}, this star is assigned a spectral type of O5, although it is consistent with any classification between O4 and O5.5. This star also has exceptionally strong \trans{he}{ii}{4686} emission and so is assigned a LC of I. Due to the strong \trans{he}{ii}{4686} emission and evidence of possible \ion{N}{3} emission, we assign a classifier of f. \citet{Evans2007} classified this star (EBU 48) as a Late O If, but from the spectrum in our sample, it is evident that it is an early O star. 

\underline{\textbf{Star 2  (O7.5 II):}} Star 2 shows evidence of weak but significant \heii\ absorption and so is classified as an O star. Based on the ratio of \trans{he}{i}{4471}/\trans{he}{ii}{4542}, this star is assigned a spectral type of O7.5, but is consistent with classifications between O7 and O8. The \trans{he}{i}{4388}/\trans{he}{ii}{4542} and \trans{he}{i}{4388}/\trans{he}{ii}{4542} ratios fall into the range appropriate for stars earlier than O8 according to \citet{Martins2018}, but these criteria are not calibrated on such early stars. It is assigned a LC of II because \trans{he}{ii}{4686} exhibits strong absorption. This star was not included in the \citet{Evans2007} sample and so we present the first published spectral type for this star.

\underline{\textbf{Star 3  (O7.5 Ia):}} Star 3 shows evidence of moderate and significant \heii\ absorption and so is classified as an O star. Based on the ratio of \trans{he}{i}{4471}/\trans{he}{ii}{4542}, this star is assigned a spectral type of O7.5. The \trans{he}{i}{4388}/\trans{he}{ii}{4542} and \trans{he}{i}{4388}/\trans{he}{ii}{4542} ratios fall into the ranges appropriate for O8-O9.7 stars according to \citet{Martins2018}, but these lines are weaker and noisier and these classifications are not calibrated for stars earlier than O8. Star 3 is assigned a LC of Ia because the \trans{he}{ii}{4686} line shows evidence of emission above the continuum. \citet{Evans2007} assigned this star (EBU 43) a spectral type of O9 If. We believe the strong \trans{he}{ii}{4542} absorption is evidence that this star is of an earlier type than O9 and we do not see evidence of \ion{N}{3} emission and so do not assign a qualifier.

\underline{\textbf{Star 4 (O8 Ib(f)):}} Star 4 shows evidence of moderate and significant \heii\ absorption and so is classified as an O star. Based on the ratio of \trans{he}{i}{4471}/\trans{he}{ii}{4542}, this star is assigned a spectral type of O8, which is supported by the \trans{he}{i}{4388}/\trans{he}{ii}{4542} and \trans{he}{i}{4388}/\trans{he}{ii}{4542} ratios. It is assigned a LC of Ib because \trans{he}{ii}{4686}/\trans{he}{i}{4713} is near neutral. It is also assigned a qualifier (f) because it shows evidence of moderate/weak \ion{N}{3} emission and neutral \trans{he}{ii}{4686}. \citet{Evans2007} assigned this star (EBU 34) a spectral type of O8 I(f) in agreement with our classification.

\begin{table*}[t]
    \caption{HST Photometry of observed stars}\label{tab:phot}
    \hspace{5in}
    \centering
\begin{tabular}{ccccccccccc}
\hline\hline
Star ID	& F225W$_\textrm{VEGA}$ 	&F275W$_\textrm{VEGA}$ 	&F336W$_\textrm{VEGA}$ 	&F475W$_\textrm{VEGA}$ 	&F814W$_\textrm{VEGA}$ \\
	& (mag) 	&(mag) 	&(mag) 	&(mag) 	&(mag)\\
(1) & (2) & (3) & (4) & (5) & (6)\\
\hline
1	& $17.507\pm0.004$ 	&$17.769\pm0.004$ 	&$18.245\pm0.004$ 	&$19.985\pm0.005$ 	&$20.192\pm0.011$ \\
2	& $17.707\pm0.004$ 	&$17.972\pm0.004$ 	&$18.481\pm0.004$ 	&$20.141\pm0.005$ 	&$20.456\pm0.013$ \\
3	& $17.272\pm0.003$ 	&$17.485\pm0.003$ 	&$17.910\pm0.003$ 	&$19.510\pm0.004$ 	&$19.669\pm0.008$ \\
4	& $17.215\pm0.003$ 	&$17.461\pm0.003$ 	&$17.906\pm0.003$ 	&$19.573\pm0.004$ 	&$19.819\pm0.009$ \\
5	& $17.441\pm0.004$ 	&$17.732\pm0.004$ 	&$18.273\pm0.004$ 	&$19.968\pm0.005$ 	&$20.253\pm0.012$ \\
6	& $17.273\pm0.003$ 	&$17.403\pm0.003$ 	&$17.788\pm0.003$ 	&$19.330\pm0.004$ 	&$19.450\pm0.008$ \\
7	& $17.599\pm0.004$ 	&$17.869\pm0.004$ 	&$18.355\pm0.004$ 	&$19.977\pm0.005$ 	&$20.266\pm0.011$ \\
8	& $17.596\pm0.004$ 	&$17.821\pm0.004$ 	&$18.304\pm0.004$ 	&$19.911\pm0.005$ 	&$20.154\pm0.011$ \\
9	& $16.528\pm0.002$ 	&$16.757\pm0.002$ 	&$17.143\pm0.002$ 	&$18.742\pm0.003$ 	&$18.896\pm0.006$ \\
10	& $17.530\pm0.004$ 	&$17.752\pm0.004$ 	&$18.190\pm0.004$ 	&$19.793\pm0.004$ 	&$19.994\pm0.010$ \\
11	& $18.127\pm0.005$ 	&$18.345\pm0.005$ 	&$18.875\pm0.005$ 	&$20.430\pm0.006$ 	&$20.688\pm0.020$ \\
12	& $18.272\pm0.008$ 	&$18.420\pm0.007$ 	&$18.959\pm0.008$ 	&$20.267\pm0.009$ 	&$20.426\pm0.019$ \\
13	& $18.056\pm0.005$ 	&$18.243\pm0.004$ 	&$18.720\pm0.005$ 	&$20.196\pm0.005$ 	&$20.341\pm0.012$ \\
14	& $18.201\pm0.005$ 	&$18.333\pm0.005$ 	&$18.625\pm0.005$ 	&$19.943\pm0.005$ 	&$20.050\pm0.011$ \\
15	& $18.080\pm0.005$ 	&$18.270\pm0.005$ 	&$18.661\pm0.005$ 	&$20.170\pm0.005$ 	&$20.301\pm0.011$ \\
16	& $18.107\pm0.005$ 	&$18.270\pm0.005$ 	&$18.658\pm0.004$ 	&$20.083\pm0.005$ 	&$20.254\pm0.011$ \\
17	& $17.437\pm0.003$ 	&$17.595\pm0.003$ 	&$17.927\pm0.003$ 	&$19.390\pm0.004$ 	&$19.465\pm0.008$ \\
18	& $17.830\pm0.004$ 	&$18.037\pm0.004$ 	&$18.508\pm0.004$ 	&$20.146\pm0.005$ 	&$20.265\pm0.011$ \\
19	& $18.228\pm0.007$ 	&$18.413\pm0.007$ 	&$18.836\pm0.007$ 	&$20.450\pm0.009$ 	&$20.476\pm0.018$ \\
20	& $17.756\pm0.007$ 	&$18.200\pm0.006$ 	&$18.698\pm0.006$ 	&$20.377\pm0.008$ 	&$20.670\pm0.019$ \\
21	& $17.882\pm0.006$ 	&$18.076\pm0.006$ 	&$18.618\pm0.006$ 	&$20.237\pm0.008$ 	&$20.498\pm0.018$ \\
22	& $17.854\pm0.004$ 	&$18.055\pm0.004$ 	&$18.522\pm0.004$ 	&$20.108\pm0.005$ 	&$20.314\pm0.012$ \\
23	& $18.213\pm0.005$ 	&$18.390\pm0.005$ 	&$18.698\pm0.005$ 	&$20.030\pm0.005$ 	&$20.101\pm0.011$ \\
24	& $17.916\pm0.004$ 	&$18.213\pm0.005$ 	&$18.715\pm0.005$ 	&$20.398\pm0.006$ 	&$20.663\pm0.014$ \\
25	& $18.078\pm0.005$ 	&$18.332\pm0.005$ 	&$18.862\pm0.005$ 	&$20.554\pm0.006$ 	&$20.846\pm0.015$ \\
\end{tabular}

    \hspace{5in}
    \tablecomments{Apparent magnitude of each star in HST NUV and optical filters.}
\end{table*}

\underline{\textbf{Star 5 (O8 III):}} Star 5 shows evidence of strong and significant \heii\ absorption and so is classified as an O star. Based on the ratio of \trans{he}{i}{4471}/\trans{he}{ii}{4542}, this star is assigned a spectral type of O8, but is consistent with classifications between O7.5 and O8.5. The \trans{he}{i}{4388}/\trans{he}{ii}{4542} and \trans{he}{i}{4388}/\trans{he}{ii}{4542} ratios fall into the ranges appropriate for O8-O9.2 stars according to \citet{Martins2018}. It is assigned a LC of III because the \trans{he}{ii}{4686} absorption is strong. This star was not included in the \citet{Evans2007} sample and so we present the first published spectral type for this star.

\underline{\textbf{Star 6 (O8.5 II):}} Star 6 shows evidence of moderate and significant \heii\ absorption and so is classified as an O star. Based on the ratio of \trans{he}{i}{4471}/\trans{he}{ii}{4542}, this star is assigned a spectral type of O8.5, which is in general agreement with the \trans{he}{i}{4388}/\trans{he}{ii}{4542} and \trans{he}{i}{4388}/\trans{he}{ii}{4542} ratios, which fall into the ranges appropriate for O8-O9.5 stars according to \citet{Martins2018}. It is assigned a LC of II because the \trans{he}{ii}{4686} absorption is weak but not neutral. \citet{Evans2007} assigned this star (EBU 20) a spectral type of O8 I.

\underline{\textbf{Star 7 (O9 Ib):}} Star 7 shows evidence of weak but significant \heii\ absorption and so is classified as an O star. Based on the ratio of \trans{he}{i}{4471}/\trans{he}{ii}{4542}, this star is assigned a spectral type of O9. The \trans{he}{i}{4388}/\trans{he}{ii}{4542} and \trans{he}{i}{4388}/\trans{he}{ii}{4542} ratios fall into the ranges appropriate for O8 stars according to \citet{Martins2018}, but these lines are weaker and noisier. It is assigned a LC of Ib because the ratio of \trans{he}{ii}{4686}/\trans{he}{i}{4713} is approximately one. \citet{Evans2007} assigned this star (EBU 49) a spectral type of O9 II.

\underline{\textbf{Star 8 (O9 I):}} Star 8 shows evidence of moderate and significant \heii\ absorption and so is classified as an O star. Based on the ratio of \trans{he}{i}{4471}/\trans{he}{ii}{4542}, this star is assigned a spectral type of O9, which is in general agreement with the \trans{he}{i}{4388}/\trans{he}{ii}{4542} and \trans{he}{i}{4388}/\trans{he}{ii}{4542} ratios, which fall into the ranges appropriate for O8-O9 stars according to \citet{Martins2018}. It is assigned a LC of I because the ratio of \trans{he}{ii}{4686}/\trans{he}{i}{4713} is less than one. \citet{Evans2007} assigned this star (EBU 43) a spectral type of O9.5 I.

\underline{\textbf{Star 9 (O9.5 Ia):}} The spectrum of Star 9 has $\langle$S/N$\rangle \approx 45$, the highest of all the spectra in our sample. It shows evidence of weak but significant \heii\ absorption and so is classified as an O star. Based on the ratio of \trans{he}{i}{4471}/\trans{he}{ii}{4542}, this star is assigned a spectral type of O9.5, which is supported by the \trans{he}{i}{4388}/\trans{he}{ii}{4542} ratio. It is assigned a LC of Ia because the ratio \trans{he}{ii}{4686}/\trans{he}{i}{4713} is approximately 0. \citet{Evans2007} classified this star (EBU 07) as B0-1 Ia, but commented that the spectral type was uncertain due to what appeared to be very weak \heii\ 4542. We detect \heii\ absorption with significance and are confident that this source is an O star. 

\underline{\textbf{Star 10 (O9.5 III):}} Star 10 shows evidence of weak but significant \heii\ absorption and so is classified as an O star. Based on the ratio of \trans{he}{i}{4471}/\trans{he}{ii}{4542}, this star is assigned a spectral type of O9.5, which is supported by the \trans{he}{i}{4388}/\trans{he}{ii}{4542} ratio. We assign a LC of III because \trans{he}{ii}{4686} $>$ \trans{he}{i}{4713}. \citet{Evans2007} classified this star (EBU 42) as B0-2, but with our higher resolution spectra, we detect significant \heii\ absorption and so feel confident in its classification as an O star.

\underline{\textbf{Star 11 (B0-1 V):}} Star 11 shows no evidence of \heii\ absorption and so is classified as a B star. Its Balmer lines appear broader than the giant standard stars but narrower than the dwarf standard stars. They do appear significantly broader than the rest of the B-star Balmer lines in our sample, so we classify this star as a dwarf. Of the remaining lines that are relevant for B-star classification, there is only evidence of \ion{Si}{4} 4089 absorption, although it is not statistically significant. We therefore assign this star a spectral type of B0-1. This star was not included in the \citet{Evans2007} sample and so we present the first published spectral type for this star.

\underline{\textbf{Star 12 (Early B I):}} Star 12 shows no evidence of \heii\ absorption and so is classified as a B star. Its Balmer lines appear narrow as compared to the standard stars and so it is classified as a giant. The remaining lines that are relevant for B-star classification (\ion{Si}{4} 4089 and 4116; \ion{Si}{3} 4552; \ion{He}{2} 4542 and 4686; and \ion{Mg}{2} 4481) are not detected in the spectrum and so we are only able to assign a spectral type of Early B. This star was not included in the \citet{Evans2007} sample and so we present the first published spectral type for this star.

\underline{\textbf{Star 13 (Early B I):}} Star 13 shows no evidence of \heii\ absorption and so is classified as a B star. Its Balmer lines appear narrow as compared to the standard stars and so it is classified as a giant. The remaining lines that are relevant for B-star classification (\ion{Si}{4} 4089 and 4116; \ion{Si}{3} 4552; \ion{He}{2} 4542 and 4686; and \ion{Mg}{2} 4481) are not detected in the spectrum and so we are only able to assign a spectral type of Early B. 
\citet{Evans2007} assigned this star (EBU 60) a spectral type of B0-0.5.

\underline{\textbf{Star 14 (Early B I):}} Star 14 shows no evidence of \heii\ absorption and so is classified as a B star. Its Balmer lines appear narrow as compared to the standard stars and so it is classified as a giant. The remaining lines that are relevant for B-star classification (\ion{Si}{4} 4089 and 4116; \ion{Si}{3} 4552; \ion{He}{2} 4542 and 4686; and \ion{Mg}{2} 4481) are not detected in the spectrum and so we are only able to assign a spectral type of Early B. 
\citet{Evans2007} assigned this star (EBU 44) a spectral type of Early B Ib.

\underline{\textbf{Star 15 (Early B I):}} Star 15 shows no evidence of \heii\ absorption and so is classified as a B star. Its Balmer lines appear narrow as compared to the standard stars and so it is classified as a giant. The remaining lines that are relevant for B-star classification (\ion{Si}{4} 4089 and 4116; \ion{Si}{3} 4552; \ion{He}{2} 4542 and 4686; and \ion{Mg}{2} 4481) are not detected in the spectrum and so we are only able to assign a spectral type of Early B. 
 \citet{Evans2007} assigned this star (EBU 61) a spectral type of Early B.

\underline{\textbf{Star 16 (Early B I):}} Star 16 shows no evidence of \heii\ absorption and so is classified as a B star. Its Balmer lines appear narrow as compared to the standard stars and so it is classified as a giant. The remaining lines that are relevant for B-star classification (\ion{Si}{4} 4089 and 4116; \ion{Si}{3} 4552; \ion{He}{2} 4542 and 4686; and \ion{Mg}{2} 4481) are not detected in the spectrum and so we are only able to assign a spectral type of Early B. This star was not included in the \citet{Evans2007} sample and so we present the first published spectral type for this star.

\underline{\textbf{Star 17 (B2 I):}} Star 17 shows no evidence of \heii\ absorption and so is classified as a B star. Its Balmer lines appear narrow as compared to the standard stars and so it is classified as a giant. The spectrum shows evidence of  \ion{Si}{3} 4553 absorption and no other relevant lines so we assign a spectral type of B2. \citet{Evans2007} assigned this star (EBU 22) a spectral type of B1 Ia.

\begin{table} [htbp]
    \centering
    \caption{\hei\ and \heii\ EWs used in O-star classification}\label{tab:ews}
    \hspace{5in}
\begin{tabular}{cccccccc}
\hline\hline
     StarID & \hei{} 4144&	\hei{} 4388&	\hei{} 4471&	\hei{} 4713&	\heii{} 4200&	\heii{} 4542&	\heii{} 4686\\
     & (\AA)&	(\AA)&	(\AA)&	(\AA)&	(\AA)&	(\AA)&	(\AA)\\
\hline
1&	$0.18\pm 0.14$ &	$0.22\pm 0.09$ &	$0.23\pm 0.09$ &	$0.25\pm 0.08$ &	$0.64\pm 0.10$ &	$0.82\pm 0.08$ &	$-1.73\pm 0.11$\\
2&	$0.03\pm 0.12$ &	$0.21\pm 0.09$ &	$0.47\pm 0.08$ &	$-0.13\pm 0.08$ &	$0.42\pm 0.11$ &	$0.42\pm 0.09$ &	$0.37\pm 0.08$ \\
3&	$0.40\pm 0.12$ &	$0.42\pm 0.08$ &	$0.84\pm 0.08$ &	$0.28\pm 0.06$ &	$0.32\pm 0.08$ &	$0.72\pm 0.07$ &	$-0.26\pm 0.07$\\
4&	$0.24\pm 0.15$ &	$-0.06\pm 0.10$ &	$0.79\pm 0.08$ &	$0.35\pm 0.07$ &	$0.58\pm 0.13$ &	$0.50\pm 0.09$ &	$-0.05\pm 0.08$\\
5&	$0.22\pm 0.14$ &	$-0.01\pm 0.09$ &	$0.69\pm 0.09$ &	$0.15\pm 0.08$ &	$0.36\pm 0.12$ &	$0.51\pm 0.08$ &	$0.63\pm 0.08$ \\
6&	$0.76\pm 0.09$ &	$0.16\pm 0.04$ &	$0.57\pm 0.05$ &	$0.24\pm 0.05$ &	$0.49\pm 0.08$ &	$0.35\pm 0.05$ &	$0.17\pm 0.04$ \\
7&	$0.00\pm 0.13$ &	$0.19\pm 0.09$ &	$0.89\pm 0.08$ &	$0.31\pm 0.07$ &	$0.26\pm 0.11$ &	$0.40\pm 0.07$ &	$0.24\pm 0.07$ \\
8&	$0.21\pm 0.16$ &	$0.24\pm 0.09$ &	$1.07\pm 0.10$ &	$0.37\pm 0.09$ &	$0.22\pm 0.14$ &	$0.40\pm 0.10$ &	$0.26\pm 0.09$ \\
9&	$0.05\pm 0.07$ &	$0.39\pm 0.04$ &	$0.74\pm 0.04$ &	$0.29\pm 0.03$ &	$0.33\pm 0.06$ &	$0.20\pm 0.04$ &	$0.04\pm 0.04$ \\
10&	$0.22\pm 0.18$ &	$0.29\pm 0.11$ &	$1.03\pm 0.10$ &	$0.11\pm 0.09$ &	$-0.23\pm 0.15$ &	$0.28\pm 0.10$ &	$0.30\pm 0.09$ 
\end{tabular}
    \hspace{5in}
\end{table}

\section{Model fitting to optical spectroscopy for all stars}\label{app:allfits}

\begin{figure*}[t]
\begin{center}
\includegraphics[width=\linewidth,angle=0]{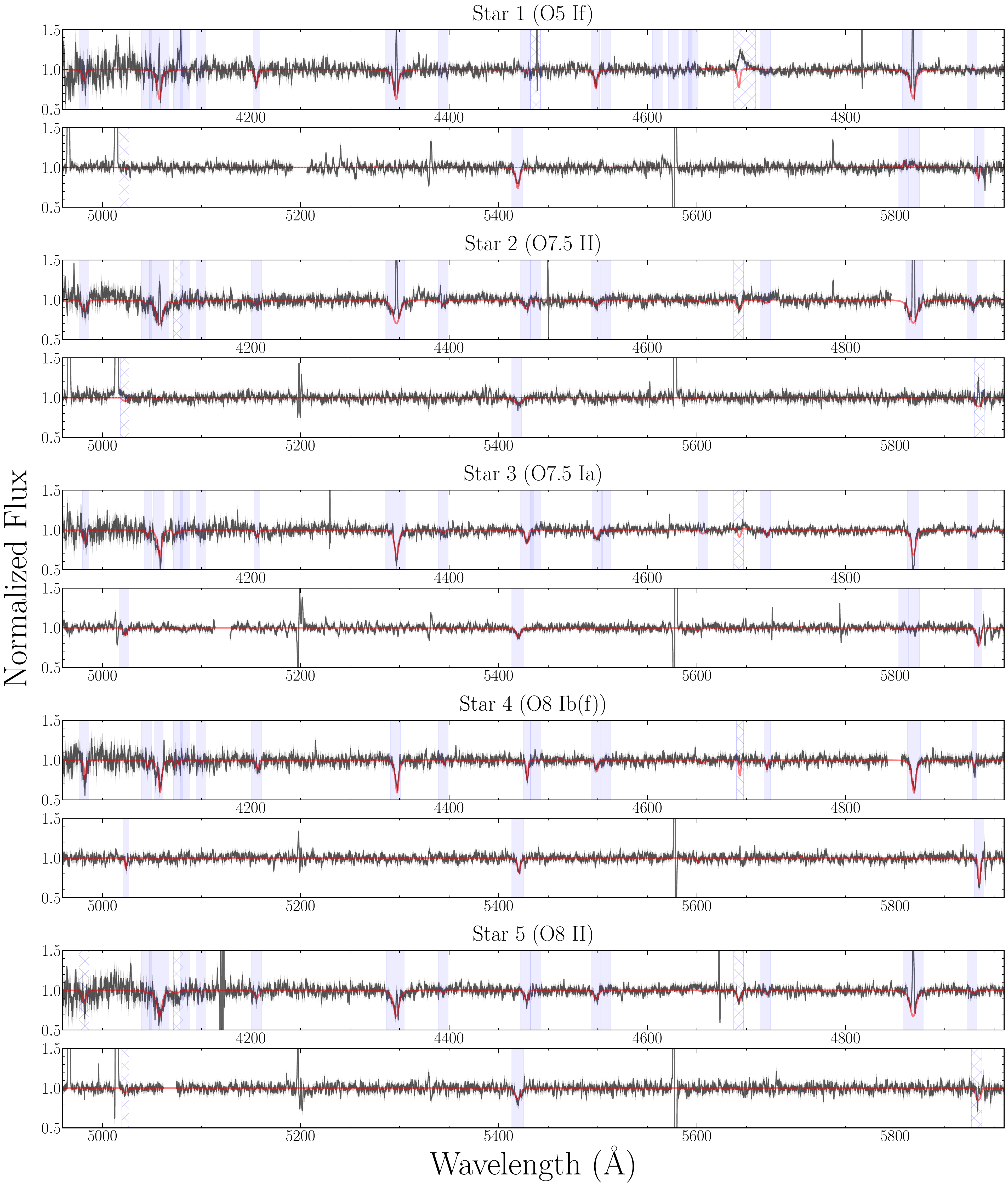}
\caption{The PoWR model atmosphere fit to the optical spectrum of Stars 1--5. The black line is the stellar spectrum, the gray shaded region is the one $\sigma$ error on the spectrum, and the red line is the PoWR model spectrum. The blue shaded regions represent lines that are included in the spectral fitting and the blue hatched regions are lines that are excluded from the fitting. \label{fig:1-5}}
\end{center}
\end{figure*}

\begin{figure*}[t]
\begin{center}
\includegraphics[width=1.01\linewidth,angle=0]{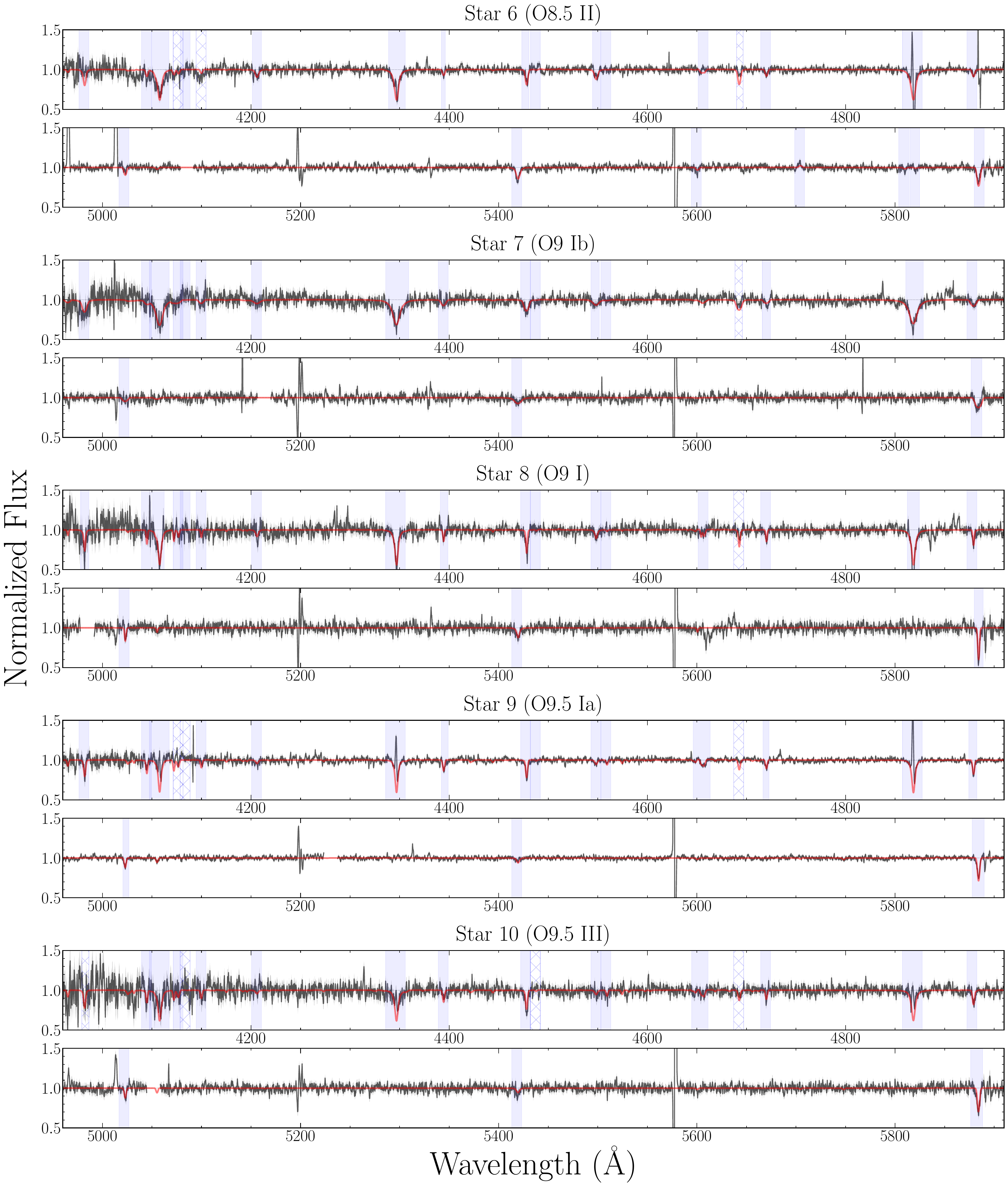}
\caption{Same as \autoref{fig:1-5}, but for Stars 6--10.\label{fig:6-10}}
\end{center}
\end{figure*}

\begin{figure*}[t]
\begin{center}
\includegraphics[width=1.01\linewidth,angle=0]{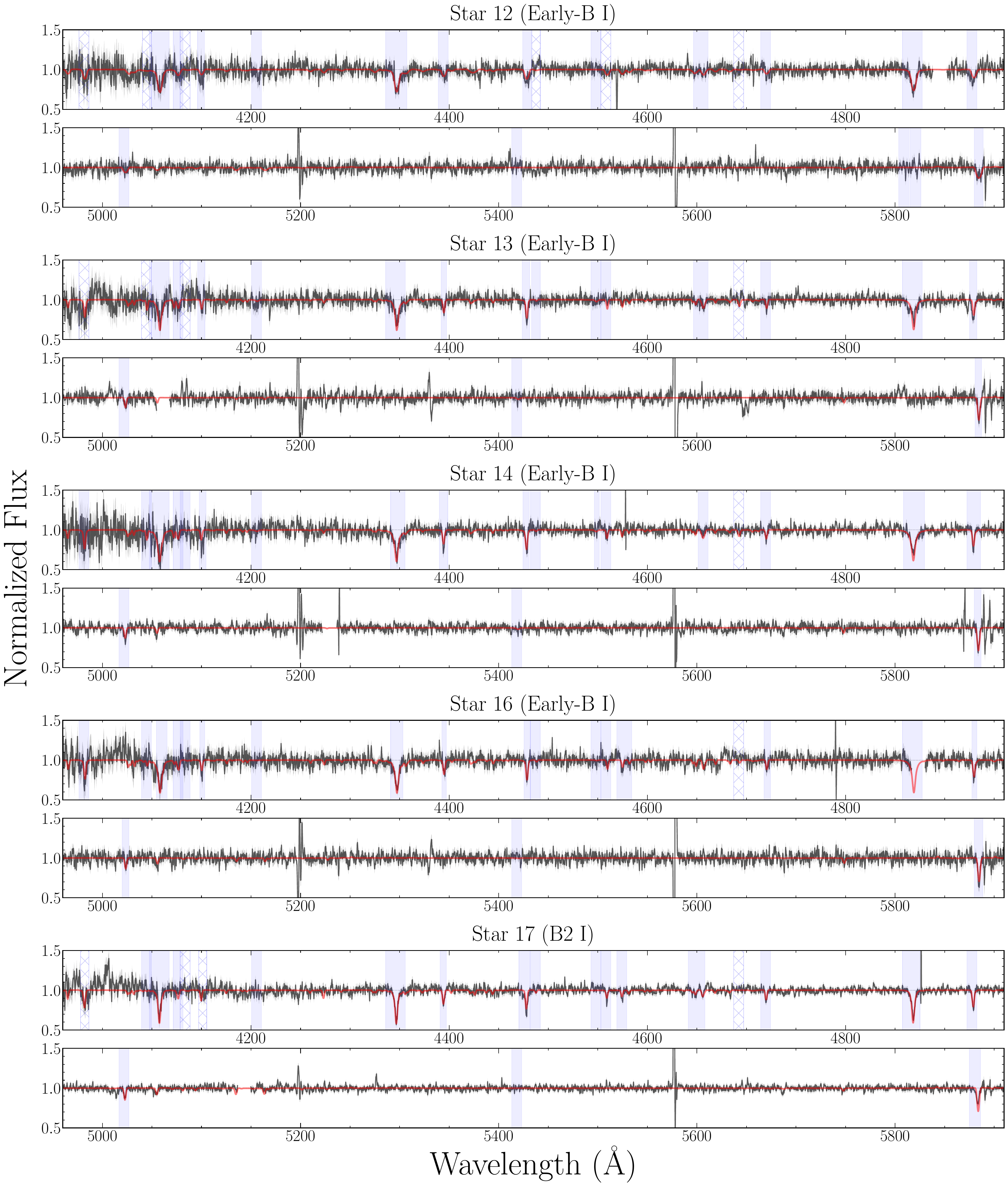}
\caption{Same as \autoref{fig:1-5}, but for Stars 12, 13, 14, 16, and 17. \label{fig:12-17}}
\end{center}
\end{figure*}

We present figures showing the optical spectroscopy and the PoWR model that provides the best fit to the spectroscopy for each star in the sample in Figures~\ref{fig:1-5}, \ref{fig:6-10}, and \ref{fig:12-17}. The lines used to fit each spectrum are highlighted in blue and are subsets of the following lines: C IV 5801, C IV 5812, H$\beta$ 4861, H$\delta$ 4102, H$\gamma$ 4340, O III 5592, \ion{C}{3} 4634, \ion{C}{3} 4640, \ion{C}{3} 4650, \ion{C}{3} 5696, \ion{C}{5} 4604, \ion{C}{5} 4620, \ion{He}{1} 4026, \ion{He}{1} 4121, \ion{He}{1} 4144, \ion{He}{1} 4388, \ion{He}{1} 4471, \ion{He}{1} 4713, \ion{He}{1} 4921, \ion{He}{1} 5015, \ion{He}{1} 5877, \ion{He}{2} 4200, \ion{He}{2} 4542, \ion{He}{2} 5411, \ion{Mg}{2} 4481, \ion{Si}{2} 4128, \ion{Si}{3} 4552, \ion{Si}{3} 4568, \ion{Si}{4} 4089.

\section{Fitted parameters to all stars for both PoWR grids}\label{app:powr}

In \autoref{tab:stellarparams} in Section~\ref{subsec:fitting_spec} we present the parameters of the PoWR models that are best-fit to the optical spectra of the OB stars in our sample. In that table, we report the parameters from the default mass-loss grid SMC-OB-VD3 for all but two stars, Stars 1 and 3, for which we report the parameters from the high mass-loss grid SMC-OB-II. We report the high mass-loss grid values for those two stars because spectral evidence of strong stellar winds -- namely \heii\ 4686 emission -- led us to believe it was more appropriate. For completeness, we also present the full set of best-fit parameters to all stars from both the default and high mass-loss grids in Table \ref{tab:powrboth}. 

In general, the fits using the high mass-loss grid prefer slightly higher temperatures and therefore slightly higher luminosities, but this trend does not hold for every star and nearly all of the best-fit parameters are statistically consistent across the two grids. 

\begin{sidewaystable}[ht]
    \caption{Best-fit parameters for all stars from both PoWR grids: SMC-OB-VD3 (default mass loss) and SMC-OB-II (high mass loss). \label{tab:powrboth}}
    \centering
{\scriptsize
\begin{tabular}{ccccccccccccccccc}
\hline\hline
     \multicolumn{1}{c}{ }&\multicolumn{8}{c}{SMC-OB-VD3 Grid} &\multicolumn{8}{c}{SMC-OB-II Grid}\\
     \cmidrule(lr){2-9}\cmidrule(lr){10-17}
     Star ID &
     T & $\log g$ & $v\sin i$& $E(B-V)_\text{int}$ & c                    & $\log(L_\star)$          &    R$_\star$               & M$_{\star,\text{spec}}$ & 
T & $\log g$ & $v\sin i$ & $E(B-V)_\text{int}$ & c                    & $\log(L_\star)$          &    R$_\star$               & M$_{\star,\text{spec}}$ \\ 
      & (kK)      & (cm s$^{-2}$) & (km s$^{-1}$) &                    &                     &   ($L_\odot$)          &    (R$_\odot$)               & (M$_\odot$)
      & (kK)      & (cm s$^{-2}$) &  (km s$^{-1}$) &                    &                     &   ($L_\odot$)          &    (R$_\odot$)               & (M$_\odot$) \\  
 \hline\\   
1 & 	39$_{-1}^{+7}$ & 	3.6$_{-0.0}^{+0.4}$ & 	160$_{-120}^{+40}$ & 	0.000$_{-0.000}^{+0.015}$ & 	0.5$_{-0.0}^{+1.1}$ & 	5.66$_{-0.04}^{+0.21}$ & 	15$_{-2}^{+1}$ & 	32$_{-0}^{+34}$  & 	45$_{-3}^{+1}$ & 	4.0$_{-0.2}^{+0.2}$ & 	80$_{-80}^{+80}$ & 	0.015$_{-0.010}^{+0.005}$ & 	1.4$_{-0.6}^{+1.8}$ & 	5.84$_{-0.10}^{+0.03}$ & 	14$_{-1}^{+1}$ & 	68$_{-23}^{+44}$ \\
2 & 	34$_{-2}^{+2}$ & 	4.0$_{-0.2}^{+0.2}$ & 	280$_{-40}^{+120}$ & 	0.000$_{-0.000}^{+0.005}$ & 	4.0$_{-2.2}^{+4.7}$ & 	5.40$_{-0.06}^{+0.07}$ & 	14$_{-1}^{+1}$ & 	75$_{-31}^{+54}$  & 	35$_{-1}^{+2}$ & 	4.2$_{-0.2}^{+0.0}$ & 	320$_{-80}^{+80}$ & 	0.005$_{-0.005}^{+0.005}$ & 	6.4$_{-2.4}^{+1.1}$ & 	5.44$_{-0.05}^{+0.07}$ & 	14$_{-1}^{+1}$ & 	119$_{-43}^{+8}$ \\
3 & 	31$_{-1}^{+1}$ & 	3.2$_{-0.0}^{+0.2}$ & 	160$_{-40}^{+40}$ & 	0.015$_{-0.000}^{+0.015}$ & 	0.7$_{-0.0}^{+1.3}$ & 	5.58$_{-0.04}^{+0.04}$ & 	21$_{-1}^{+1}$ & 	27$_{-0}^{+20}$  & 	32$_{-1}^{+2}$ & 	3.4$_{-0.2}^{+0.2}$ & 	160$_{-40}^{+40}$ & 	0.030$_{-0.015}^{+0.010}$ & 	1.2$_{-0.7}^{+2.1}$ & 	5.64$_{-0.06}^{+0.08}$ & 	21$_{-1}^{+1}$ & 	42$_{-16}^{+31}$ \\
4 & 	32$_{-1}^{+1}$ & 	3.4$_{-0.2}^{+0.2}$ & 	80$_{-40}^{+40}$ & 	0.000$_{-0.000}^{+0.010}$ & 	1.2$_{-0.7}^{+1.7}$ & 	5.56$_{-0.03}^{+0.05}$ & 	20$_{-1}^{+1}$ & 	35$_{-11}^{+26}$  & 	32$_{-1}^{+1}$ & 	3.4$_{-0.2}^{+0.2}$ & 	80$_{-40}^{+40}$ & 	0.000$_{-0.000}^{+0.010}$ & 	1.0$_{-0.6}^{+1.8}$ & 	5.57$_{-0.04}^{+0.05}$ & 	20$_{-1}^{+1}$ & 	36$_{-12}^{+26}$ \\
5 & 	35$_{-2}^{+1}$ & 	3.8$_{-0.2}^{+0.2}$ & 	200$_{-80}^{+40}$ & 	0.000$_{-0.000}^{+0.000}$ & 	2.3$_{-1.3}^{+2.7}$ & 	5.52$_{-0.06}^{+0.06}$ & 	16$_{-1}^{+1}$ & 	56$_{-21}^{+37}$  & 	35$_{-1}^{+3}$ & 	3.8$_{-0.2}^{+0.4}$ & 	200$_{-40}^{+80}$ & 	0.000$_{-0.000}^{+0.000}$ & 	2.2$_{-1.4}^{+3.7}$ & 	5.52$_{-0.04}^{+0.09}$ & 	16$_{-2}^{+1}$ & 	56$_{-21}^{+71}$ \\
6 & 	33$_{-1}^{+2}$ & 	3.6$_{-0.2}^{+0.2}$ & 	120$_{-40}^{+40}$ & 	0.075$_{-0.010}^{+0.010}$ & 	3.5$_{-2.3}^{+4.4}$ & 	5.80$_{-0.05}^{+0.08}$ & 	24$_{-1}^{+2}$ & 	85$_{-33}^{+64}$  & 	34$_{-1}^{+2}$ & 	3.6$_{-0.2}^{+0.4}$ & 	160$_{-40}^{+40}$ & 	0.080$_{-0.015}^{+0.010}$ & 	2.8$_{-1.9}^{+7.2}$ & 	5.85$_{-0.06}^{+0.07}$ & 	24$_{-1}^{+1}$ & 	85$_{-34}^{+123}$ \\
7 & 	34$_{-2}^{+1}$ & 	4.2$_{-0.2}^{+0.0}$ & 	240$_{-40}^{+40}$ & 	0.015$_{-0.010}^{+0.000}$ & 	9.2$_{-4.9}^{+1.1}$ & 	5.48$_{-0.08}^{+0.03}$ & 	16$_{-1}^{+1}$ & 	146$_{-61}^{+7}$  & 	34$_{-0}^{+2}$ & 	4.2$_{-0.0}^{+0.0}$ & 	240$_{-40}^{+40}$ & 	0.015$_{-0.000}^{+0.005}$ & 	8.7$_{-1.9}^{+0.0}$ & 	5.49$_{-0.00}^{+0.06}$ & 	16$_{-1}^{+0}$ & 	149$_{-14}^{+0}$ \\
8 & 	31$_{-2}^{+1}$ & 	3.4$_{-0.2}^{+0.2}$ & 	40$_{-40}^{+40}$ & 	0.020$_{-0.015}^{+0.005}$ & 	1.0$_{-0.7}^{+1.6}$ & 	5.41$_{-0.09}^{+0.04}$ & 	18$_{-1}^{+1}$ & 	28$_{-11}^{+21}$  & 	34$_{-1}^{+1}$ & 	3.8$_{-0.2}^{+0.2}$ & 	40$_{-40}^{+40}$ & 	0.035$_{-0.010}^{+0.005}$ & 	2.7$_{-1.6}^{+3.3}$ & 	5.52$_{-0.05}^{+0.04}$ & 	17$_{-1}^{+1}$ & 	64$_{-26}^{+44}$ \\
9 & 	27$_{-1}^{+1}$ & 	3.0$_{-0.0}^{+0.2}$ & 	80$_{-40}^{+40}$ & 	0.005$_{-0.005}^{+0.010}$ & 	1.1$_{-0.3}^{+2.3}$ & 	5.74$_{-0.05}^{+0.05}$ & 	34$_{-2}^{+2}$ & 	42$_{-4}^{+30}$  & 	27$_{-1}^{+1}$ & 	3.0$_{-0.0}^{+0.2}$ & 	120$_{-40}^{+40}$ & 	0.005$_{-0.005}^{+0.015}$ & 	1.0$_{-0.3}^{+2.2}$ & 	5.75$_{-0.04}^{+0.07}$ & 	34$_{-1}^{+2}$ & 	43$_{-2}^{+33}$ \\
10 & 	28$_{-2}^{+1}$ & 	3.2$_{-0.2}^{+0.2}$ & 	80$_{-40}^{+40}$ & 	0.005$_{-0.005}^{+0.010}$ & 	0.9$_{-0.6}^{+1.5}$ & 	5.36$_{-0.07}^{+0.06}$ & 	20$_{-1}^{+2}$ & 	24$_{-9}^{+18}$  & 	29$_{-2}^{+1}$ & 	3.2$_{-0.2}^{+0.2}$ & 	80$_{-40}^{+40}$ & 	0.010$_{-0.010}^{+0.010}$ & 	0.6$_{-0.4}^{+1.2}$ & 	5.40$_{-0.08}^{+0.05}$ & 	20$_{-1}^{+2}$ & 	23$_{-8}^{+16}$ \\
12 & 	20$_{-3}^{+4}$ & 	2.6$_{-0.2}^{+0.2}$ & 	200$_{-40}^{+40}$ & 	0.000$_{-0.000}^{+0.035}$ & 	0.3$_{-0.2}^{+0.5}$ & 	4.80$_{-0.08}^{+0.21}$ & 	21$_{-3}^{+6}$ & 	6$_{-3}^{+6}$  & 	17$_{-1}^{+2}$ & 	2.4$_{-0.0}^{+0.2}$ & 	240$_{-40}^{+40}$ & 	0.000$_{-0.000}^{+0.000}$ & 	0.3$_{-0.1}^{+0.6}$ & 	4.72$_{-0.03}^{+0.06}$ & 	26$_{-4}^{+4}$ & 	6$_{-2}^{+7}$ \\
13 & 	25$_{-1}^{+1}$ & 	3.0$_{-0.2}^{+0.2}$ & 	80$_{-40}^{+40}$ & 	0.010$_{-0.010}^{+0.010}$ & 	0.5$_{-0.4}^{+0.8}$ & 	5.08$_{-0.07}^{+0.06}$ & 	19$_{-1}^{+1}$ & 	13$_{-5}^{+9}$  & 	28$_{-1}^{+1}$ & 	3.2$_{-0.2}^{+0.4}$ & 	120$_{-40}^{+40}$ & 	0.035$_{-0.010}^{+0.010}$ & 	0.6$_{-0.4}^{+1.8}$ & 	5.22$_{-0.06}^{+0.06}$ & 	17$_{-1}^{+1}$ & 	17$_{-7}^{+27}$ \\
14 & 	26$_{-9}^{+1}$ & 	3.2$_{-0.8}^{+0.2}$ & 	80$_{-40}^{+40}$ & 	0.100$_{-0.100}^{+0.010}$ & 	1.5$_{-1.1}^{+2.1}$ & 	5.35$_{-0.50}^{+0.06}$ & 	23$_{-1}^{+8}$ & 	31$_{-23}^{+21}$  & 	29$_{-13}^{+1}$ & 	3.6$_{-1.2}^{+0.2}$ & 	80$_{-40}^{+40}$ & 	0.130$_{-0.130}^{+0.005}$ & 	3.9$_{-3.3}^{+4.8}$ & 	5.51$_{-0.67}^{+0.04}$ & 	22$_{-2}^{+13}$ & 	73$_{-63}^{+48}$ \\
16 & 	23$_{-4}^{+3}$ & 	3.0$_{-0.4}^{+0.2}$ & 	80$_{-40}^{+40}$ & 	0.010$_{-0.010}^{+0.040}$ & 	0.8$_{-0.6}^{+1.3}$ & 	5.05$_{-0.16}^{+0.18}$ & 	21$_{-1}^{+5}$ & 	16$_{-7}^{+12}$  & 	26$_{-3}^{+1}$ & 	3.2$_{-0.2}^{+0.2}$ & 	80$_{-40}^{+40}$ & 	0.050$_{-0.035}^{+0.010}$ & 	1.0$_{-0.7}^{+1.5}$ & 	5.23$_{-0.17}^{+0.06}$ & 	20$_{-1}^{+2}$ & 	24$_{-11}^{+16}$ \\
17 & 	18$_{-1}^{+3}$ & 	2.4$_{-0.0}^{+0.2}$ & 	80$_{-40}^{+40}$ & 	0.000$_{-0.000}^{+0.000}$ & 	0.5$_{-0.2}^{+1.2}$ & 	5.14$_{-0.02}^{+0.10}$ & 	38$_{-7}^{+5}$ & 	13$_{-2}^{+14}$  & 	18$_{-1}^{+8}$ & 	2.4$_{-0.0}^{+0.6}$ & 	120$_{-40}^{+40}$ & 	0.000$_{-0.000}^{+0.060}$ & 	0.5$_{-0.2}^{+1.1}$ & 	5.15$_{-0.03}^{+0.38}$ & 	39$_{-10}^{+5}$ & 	14$_{-3}^{+19}$ 
\end{tabular}}
    \tablecomments{Columns are the same as described for \autoref{tab:stellarparams}.}
\end{sidewaystable}

\section{Fitting with TLUSTY models}\label{app:tlusty}

As discussed in Section~\ref{sec:fitting}, to fit our optical spectroscopy and NUV photometry, we rely primarily on grids of stellar atmosphere models calculated using the PoWR code \citep{Grafener2002, Hamann2003, Sander2015}. Specifically, we use the set of grids generated with SMC metallicity (20\%\,$Z_\odot$), which at the time of analysis was the lowest metallicity of the publicly available grids. Given that the gas-phase metallicity of NGC 3109 is actually lower than the SMC at 12\%\,$Z_\odot$, we investigated the impact of this choice by repeating the fitting procedure described in Section~\ref{subsec:fitting_spec} with a different set of model atmosphere grids. 

Similar to the PoWR code, TLUSTY generates non-LTE stellar atmosphere models and provides public grids of synthetic optical spectra. The TLUSTY grids are separated into O and B stars, but span a broader range of metallicity than the PoWR grids. To assess the sensitivity of the best-fit parameters to the choice of grid metallicity, we fit the stellar spectra using three TLUSTY model grids with different metallicities: 50\%\,$Z_\odot$, 20\%\,$Z_\odot$, and 10\%\,$Z_\odot$. We report the best-fit values of $T_\star$, $\log(g)$, and $v\sin i$ in \autoref{tab:tlusty}.

The fitted parameters are generally very insensitive to changes in the grid metallicity. It is possible that this is a reflection of the low SN of our spectra and the uncertainty in the stellar metallicity of each source. However, we do find evidence that the LMC metallicity grid overpredicts the strength of some metal lines and is therefore not an ideal choice for our sample. The fits to Star 9 show the largest discrepancy in fitted parameters among the grids. The spectrum of Star 9 has the highest SN of the spectra in our sample and so is best suited to distinguish between the models. We find that the optimal model from the LMC-grid is not well-fit to the spectrum of Star 9 as it contains stronger metal absorption lines than are evident in the spectrum. This is most apparent for \trans{Si}{iii}{4557} and \trans{C}{ii}{4267}.

While confirmation of the stellar abundances in these stars would require additional data, this experiment provides evidence justifying our choice of the PoWR SMC metallicity grids and demonstrates that using a different metallicity grid would have a minimal impact on our results.

\begin{table*}[ht]
    \caption{Best-fit parameters derived from fits to TLUSTY grids of varying metallicity.}\label{tab:tlusty}
    \hspace{5in}
    \centering
\begin{tabular}{cccccccccc}
\hline\hline
     \multicolumn{1}{c}{ }&\multicolumn{3}{c}{50\% $Z_\odot$} &\multicolumn{3}{c}{20\% $Z_\odot$}&\multicolumn{3}{c}{10\% $Z_\odot$}\\
     \cmidrule(lr){2-4}\cmidrule(lr){5-7}\cmidrule(lr){8-10}
     Star ID &
     T & $\log g$ & $v\sin i$ & 
     T & $\log g$ & $v\sin i$ & 
     T & $\log g$ & $v\sin i$ \\  
      & (kK)      & (cm s$^{-2}$) & (km s$^{-1}$)  & (kK)      & (cm s$^{-2}$) & (km s$^{-1}$)  & (kK)      & (cm s$^{-2}$) & (km s$^{-1}$)   \\
(1) & (2) & (3) & (4) & (5) & (6) & (7) & (8) & (9) & (10)\\
     \cmidrule(lr){1-1}\cmidrule(lr){2-4}\cmidrule(lr){5-7}\cmidrule(lr){8-10}
1 & 	50$_{-5}^{+5}$ & 	4.00$_{-0.25}^{+0.25}$ & 	80$_{-80}^{+40}$  & 	50$_{-5}^{+5}$ & 	4.00$_{-0.25}^{+0.25}$ & 	80$_{-80}^{+40}$  & 	50$_{-5}^{+5}$ & 	4.00$_{-0.25}^{+0.25}$ & 	80$_{-80}^{+40}$ \\
2 & 	35$_{-2}^{+2}$ & 	4.25$_{-0.50}^{+0.25}$ & 	280$_{-40}^{+120}$  & 	35$_{-1}^{+2}$ & 	4.25$_{-0.25}^{+0.25}$ & 	280$_{-40}^{+160}$  & 	35$_{-1}^{+2}$ & 	4.00$_{-0.25}^{+0.50}$ & 	320$_{-80}^{+120}$ \\
3 & 	32$_{-1}^{+2}$ & 	3.25$_{-0.25}^{+0.25}$ & 	160$_{-40}^{+40}$  & 	32$_{-1}^{+2}$ & 	3.25$_{-0.25}^{+0.25}$ & 	160$_{-40}^{+40}$  & 	32$_{-1}^{+2}$ & 	3.25$_{-0.25}^{+0.25}$ & 	160$_{-40}^{+40}$ \\
4 & 	32$_{-1}^{+1}$ & 	3.50$_{-0.25}^{+0.25}$ & 	80$_{-40}^{+40}$  & 	32$_{-1}^{+2}$ & 	3.50$_{-0.25}^{+0.25}$ & 	80$_{-40}^{+40}$  & 	32$_{-1}^{+1}$ & 	3.25$_{-0.25}^{+0.25}$ & 	80$_{-40}^{+40}$ \\
5 & 	35$_{-1}^{+2}$ & 	3.75$_{-0.25}^{+0.50}$ & 	200$_{-80}^{+40}$  & 	38$_{-2}^{+2}$ & 	4.00$_{-0.25}^{+0.25}$ & 	200$_{-80}^{+40}$  & 	38$_{-2}^{+2}$ & 	4.00$_{-0.25}^{+0.25}$ & 	200$_{-80}^{+40}$ \\
6 & 	35$_{-1}^{+2}$ & 	3.75$_{-0.25}^{+0.25}$ & 	120$_{-40}^{+40}$  & 	35$_{-1}^{+1}$ & 	3.75$_{-0.25}^{+0.25}$ & 	120$_{-40}^{+40}$  & 	35$_{-1}^{+1}$ & 	3.75$_{-0.25}^{+0.25}$ & 	120$_{-40}^{+40}$ \\
7 & 	35$_{-2}^{+1}$ & 	4.25$_{-0.25}^{+0.25}$ & 	200$_{-40}^{+40}$  & 	35$_{-1}^{+2}$ & 	4.25$_{-0.25}^{+0.50}$ & 	200$_{-40}^{+40}$  & 	35$_{-1}^{+2}$ & 	4.25$_{-0.25}^{+0.50}$ & 	200$_{-40}^{+40}$ \\
8 & 	32$_{-1}^{+1}$ & 	3.50$_{-0.25}^{+0.25}$ & 	40$_{-40}^{+40}$  & 	32$_{-1}^{+1}$ & 	3.50$_{-0.25}^{+0.25}$ & 	40$_{-40}^{+40}$  & 	32$_{-1}^{+2}$ & 	3.50$_{-0.25}^{+0.25}$ & 	40$_{-40}^{+40}$ \\
9 & 	17$_{-2}^{+9}$ & 	2.25$_{-0.25}^{+0.50}$ & 	80$_{-40}^{+40}$  & 	26$_{-1}^{+2}$ & 	2.75$_{-0.25}^{+0.25}$ & 	80$_{-40}^{+40}$  & 	28$_{-2}^{+1}$ & 	3.00$_{-0.25}^{+0.25}$ & 	80$_{-40}^{+40}$ \\
10 & 	28$_{-1}^{+1}$ & 	3.00$_{-0.25}^{+0.25}$ & 	80$_{-40}^{+40}$  & 	28$_{-1}^{+2}$ & 	3.00$_{-0.25}^{+0.25}$ & 	80$_{-40}^{+40}$  & 	28$_{-1}^{+2}$ & 	3.00$_{-0.25}^{+0.25}$ & 	80$_{-40}^{+40}$ \\
12 & 	17$_{-2}^{+3}$ & 	2.25$_{-0.25}^{+0.25}$ & 	200$_{-40}^{+40}$  & 	20$_{-3}^{+1}$ & 	2.50$_{-0.25}^{+0.25}$ & 	200$_{-40}^{+40}$  & 	20$_{-3}^{+4}$ & 	2.50$_{-0.25}^{+0.25}$ & 	200$_{-40}^{+40}$ \\
13 & 	18$_{-1}^{+2}$ & 	2.25$_{-0.25}^{+0.25}$ & 	80$_{-40}^{+40}$  & 	20$_{-2}^{+8}$ & 	2.50$_{-0.25}^{+0.75}$ & 	80$_{-40}^{+40}$  & 	23$_{-3}^{+4}$ & 	2.75$_{-0.25}^{+0.50}$ & 	80$_{-40}^{+40}$ \\
14 & 	18$_{-1}^{+1}$ & 	2.50$_{-0.25}^{+0.25}$ & 	80$_{-40}^{+40}$  & 	19$_{-1}^{+2}$ & 	2.50$_{-0.25}^{+0.25}$ & 	80$_{-40}^{+40}$  & 	21$_{-3}^{+3}$ & 	2.75$_{-0.25}^{+0.25}$ & 	80$_{-40}^{+40}$ \\
16 & 	21$_{-3}^{+6}$ & 	2.75$_{-0.25}^{+0.50}$ & 	80$_{-40}^{+40}$  & 	21$_{-2}^{+4}$ & 	2.75$_{-0.25}^{+0.25}$ & 	40$_{-40}^{+40}$  & 	24$_{-4}^{+1}$ & 	3.00$_{-0.25}^{+0.25}$ & 	40$_{-40}^{+40}$ \\
17 & 	18$_{-1}^{+1}$ & 	2.25$_{-0.25}^{+0.25}$ & 	80$_{-40}^{+40}$  & 	18$_{-1}^{+1}$ & 	2.25$_{-0.25}^{+0.25}$ & 	80$_{-40}^{+40}$  & 	20$_{-2}^{+1}$ & 	2.50$_{-0.25}^{+0.25}$ & 	80$_{-40}^{+40}$
\end{tabular}
    \hspace{5in}
    \tablecomments{Columns (2) -- (4) are parameters derived from TLUSTY grids with 50\% $Z_\odot$ metallicity. Columns (5) -- (7) are parameters derived from TLUSTY grids with 20\% $Z_\odot$ metallicity. Columns (8) -- (10) include parameters derived from TLUSTY grids with 10\% $Z_\odot$ metallicity}
\end{table*}

\bibliography{NGC3109}
\end{document}